\documentclass[twocolumn, astrosymb]{aastex631}

\usepackage[normalem]{ulem}
\usepackage{amsmath, mathtools, esdiff}
\usepackage{xspace}
\let\tablenum\relax
\usepackage[print-unity-mantissa = false]{siunitx}

\newcommand{\dint}[1]{~\mathrm{d}#1}

\newcommand{\MJ}{M_{\mathrm{J}}}
\newcommand{\ME}{M_{\mathrm{E}}}
\DeclareSIUnit[]{\kb}{k_B}
\DeclareSIUnit[]{\yr}{yr}
\DeclareSIUnit{\bar}{bar}
\DeclareSIUnit{\bary}{m_u}
\DeclareSIUnit{\kbperbary}{\kb \per \bary}

\newcommand{\Zatm}{Z_\mathrm{atm}}
\newcommand{\Zbulk}{Z_\mathrm{bulk}}
\newcommand{\sXZ}{s_{X,Z}}
\newcommand{\dt}{\delta t}
\newcommand{\dS}{\delta S}
\newcommand{\ds}{\Delta s}
\newcommand{\dsXZ}{\Delta s_{X,Z}}

\newcommand{\mrcb}{m_\mathrm{RCB}}
\newcommand{\senv}{s_\mathrm{env}}
\newcommand{\mesa}{\texttt{MESA}\xspace}
\newcommand{\cpm}{\texttt{convective\_premixing}\xspace}
\newcommand{\gm}{\texttt{gentle\_mixing}\xspace}
\newcommand{\mh}{\texttt{py\_mesa\_helper}\xspace}

\begin{document}

\title{Convective Mixing in Gas Giant Planets with Primordial Composition Gradients}

\author[0009-0003-0637-9170]{Henrik Knierim}
\affiliation{Department of Astrophysics \\
University of Zurich\\
Winterthurerstr. 190\\
CH-8057 Zurich, Switzerland}
\correspondingauthor{Henrik Knierim}
\email{henrik.knierim@uzh.ch}

\author[0000-0001-5555-2652]{Ravit Helled}
\affiliation{Department of Astrophysics \\
University of Zurich\\
Winterthurerstr. 190\\
CH-8057 Zurich, Switzerland}

\begin{abstract}
Linking atmospheric measurements to the bulk planetary composition and ultimately the planetary origin is a key objective in planetary science.
In this work, we identify the cases in which the atmospheric composition represents the bulk composition. 
We simulate the evolution of giant planets considering a wide range of planetary masses ($0.3-\SI{2}{\MJ}$), initial entropies ($8-\SI{11}{\kbperbary}$), and primordial heavy-element profiles. 
We find that convective mixing is most efficient at early times (ages $\lesssim$ \SI{1e7}{\yr}) and that primordial composition gradients can be eroded. In several cases, however, the atmospheric composition can differ widely from the planetary bulk composition, with the exact outcome depending on the details. We show that the efficiency of convection is primarily controlled by the underlying entropy profile: For low primordial entropies of $8-\SI{9}{\kbperbary}$, convective mixing can be inhibited and composition gradients can persist over billions of years. 
The scaling of mixing efficiency with mass is governed by the primordial entropy. For the same primordial entropy, low-mass planets mix more efficiently than high-mass planets. If the primordial internal entropy would increase with mass, however, this trend could reverse.
We also present a new analytical model that predicts convective mixing under the existence of composition (and entropy) gradients.
Our results emphasize the complexity in the interpretation of atmospheric abundance measurements and show the great need to better understand the planetary formation process as it plays a key role in determining the planetary evolution and final structure. 
\end{abstract}


\section{Introduction} \label{sec:intro}
A new era in giant planet exploration has begun with the advent of accurate atmospheric measurements of giant exoplanets by space missions like JWST \citep{Gardner2006} and ARIEL \citep{Tinetti2018}. 
A key goal of exoplanetary science is to use these new data to reveal information on the planetary bulk composition and formation history. 
Several studies have suggested that determining the bulk composition of giant planets can constrain their formation process and early evolution \citep[e.g.,][]{Oeberg_2011, Madhusudhan_2017, Schneider_2021b, Turrini_2021, Hands_2022, Knierim_2022}.
However, it is still unknown how well atmospheric abundances trace the planetary bulk composition and under what conditions giant planets are expected to be fully mixed \cite[e.g.,][]{Madhusudhan_2012, Turrini_2018, Madhusudhan_2019, Helled2022_Ariel, Molliere_2022}.
Typically, when inferring the planetary bulk composition, 
most models assume a simplified internal structure (homogeneously mixed or core+envelope). However, we now know that both Jupiter and Saturn have extended ``fuzzy" cores and inhomogeneous interiors \citep[e.g.,][]{Fuller_2014, Wahl2017,mankovich2021, 2021PSJ.....2..241N,Helled2022_Jupiter, Idini_2022, Miguel2022, Dewberry_2023, 2023A&A...680L...2H}. It is therefore possible that many of the observed exoplanets also have more complex internal structures. These exoplanets typically have ages of several gigayears. Therefore, in order to link their present-day internal structures with their origins, we must understand the long-term evolution of planetary interiors.
\par
Currently, there is still a gap in our understanding of when giant planets are expected to be fully mixed. Although detailed hydrodynamic simulations of convection start to approach the conditions relevant for giant planets \citep[e.g.,][]{Radko_2007, Rosenblum_2011, Aurnou_2020, Fuentes_2020, Fuentes_2022, Tulekeyev_2024}, they are limited both in temporal and spatial scale. While some effort has been made to model the mixing of primordial composition profiles on evolutionary timescales for Jupiter \citep[e.g.,][]{Vazan2018, Mueller2020a, Helled2022_Jupiter}, a comprehensive investigation of mixing in giant planets in more general terms is still required. 
This is a challenging problem due to the uncertainty in both the primordial composition gradient and the primordial entropy. 
\par
Simulations of giant planet formation in the core-accretion framework \citep{Pollack1996} find that the post-formation heavy-element mass fraction begins at a high central value, which corresponds to a compact heavy-element core with $Z \lesssim 1$, and then decreases (continuously or sharply) toward a nearly constant heavy-element mass fraction in the envelope \citep[e.g.][]{Helled2017, Lozovsky2017, Bodenheimer_2018, Valletta2020, Stevenson2022}. The exact shape of the composition profile depends on various parameters such as the accretion rate, the accretion mechanism, the opacity of the material, and the chemical properties of the species that are being accreted (e.g., condensation temperatures). Similarly, depending on the shock physics, the planetary entropy can drastically increase or decrease during runaway gas accretion, leading to the so-called ``entropy tuning fork" \citep[e.g.][]{Marley2007, Mordasini2013, Berardo2017a, Berardo2017b, Mordasini2017, Cumming2018}. \citet{Cumming2018}, for example, showed that the primordial entropy profile of Jupiter is not adiabatic, leading to a radiative inner region for the first $\sim\SI{10}{Myr}$ of their simulations. Overall, the primordial entropy and heavy-element distribution in giant planets remain largely unknown. 

In this study, we investigate the planetary evolution with convective mixing and highlight trends and dependencies of the mixing on planetary parameters. We also provide an analytical framework that can be used to assess the efficiency of convective mixing in giant planet interiors. 
In particular, we aim to answer the questions below: 
\begin{enumerate}
\item What makes a primordial composition profile stable over evolutionary timescales?
\item How does the primordial entropy profile influence the mixing?
\item How does mixing change with planetary mass?
\item What are systems where the atmospheric heavy-element mass fraction does not represent the bulk?
\end{enumerate}

Our paper is organized as follows. Section \ref{sec:numerical_methods} describes the numerical methods to simulate convective mixing in giant planets. In Section \ref{sec:analytic_model}, we present an analytic model for giant planet mixing. Equipped with these tools, Section \ref{sec:results} presents the results of our numerical experiments. Our results are discussed and summarized in Section \ref{sec:discussion}.

\section{Numerical Methods} \label{sec:numerical_methods}
\subsection{Planetary Evolution with Convective Mixing} 
We model the long-term evolution of gas giant planets using the code Modules for Experiments in Stellar Astrophysics \citep[\mesa;][]{Paxton_2011, Paxton_2013, Paxton_2015, Paxton_2018, Paxton_2019, Jermyn_2023}. To simulate convective mixing properly, we built on the improvements presented by \citet[see also \citealt{Mueller_2020b, Mueller_2021, Muller2023}]{Mueller_2020} and extended \mesa's capabilities to calculate the planetary evolution in three major ways.

First, we implemented support for user-supplied equations of state (EoSs). \citet{Mueller_2020} have already implemented EoSs that are appropriate to model giant planets for \mesa (r10108). This includes EoSs for water, rock, and a half-half mixture thereof \citep[for further information on the EoS, see][]{Mueller_2020}. For this study, we created a new module that reads any EoS supplied by the user. This module does not modify the source code of \mesa (using \texttt{run\_stars\_extras} instead) and works with all recent releases. For this study, we employ the most recent release 24.03.1. Our code is open source and available on GitHub \citep[][]{mesa_custom_eos}.\footnote{\url{https://github.com/Henrik-Knierim/mesa_custom_EoS}}

Second, we extended \mesa's \cpm algorithm that more accurately determines convective boundaries.
Convective stability in planetary interiors is typically determined using the Schwarzschild or Ledoux criterion. The Schwarzschild criterion assesses stability based on the temperature gradient, while the Ledoux criterion accounts for both temperature and composition gradients (see Section \ref{sec:analytic_model}). In short, \cpm analyzes the convective regions by tentatively mixing adjacent cells. If the neighboring cell becomes convective after mixing, it is added to the convective region, and this process continues until stable boundaries are found \citep[for details, see][]{Paxton_2018, Paxton_2019}. This method avoids the inconsistencies of a previous approach, which relied on detecting sign changes in the criterion for convective stability between grid points (for details, see Appendix \ref{sec:improvements_to_mesa}). In \cpm, semi-convective regions (Ledoux stable but Schwarzschild unstable) are treated as unstable. 
It should be noted that the exact nature of semi-convection is uncertain and the influence of mixing on the planet’s thermodynamic structure can lead to radiative regions, as discussed in Section \ref{sec:analytic_model} and Appendix \ref{sec:analytic_mixing_details}. We modified \cpm by introducing two new modes that hold either pressure and density or pressure and total entropy of the extended convective region constant. While artificially fixing any combination of variables is incorrect, the two new modes show considerably more stability compared to the constant-temperature modes under planetary conditions.
\par
Third, we developed a new mixing algorithm called \gm. Since convection is extremely efficient in erasing compositional differences, cells that become convective can alter their composition drastically within one time step. As a consequence, their physical properties like density or temperature can also change drastically. If too many cells change at once, \mesa's solver may struggle to find a convergent model. To mitigate this issue, \gm monitors the change in composition and, if it becomes too large, reduces both the mixing efficiency and the time step. Further details on our improvements to \mesa can be found in Appendix \ref{sec:improvements_to_mesa}

\subsection{Initial Model}\label{sec:initial_model}
In all the simulations, we first create a homogeneous model with a protosolar composition \citep[taken from ][]{Lodders_2021} of the specified mass and entropy profile. Next, we relax the heavy-element mass fraction profile (via the \texttt{relax\_initial\_composition} functionality) and evolve the planet for $\SI{1e3}{\yr}$ before mixing.
An increase in heavy-element mass fraction leads to contraction, and therefore higher temperature and density. The entropy after the relaxation depends on the ratio of these changes and their respective partial derivatives (i.e., the EoS). Typically, the entropy derivative with respect to composition dominates, leading to a decrease in entropy in the region of the composition gradient. In this study, when we refer to an entropy value, we mean the pre-\texttt{relax\_initial\_composition} entropy, i.e., the (specific) entropy for a protosolar composition, unless stated otherwise.
We use heavy-element mass fraction profiles that interpolate between a constant core region, one or two exponential decays, and a constant-envelope region using cubic functions. Requiring that $X/Y = X_\mathrm{proto\_solar}/Y_\mathrm{proto\_solar}$ yields $X(m)$ and $Y(m)$, where $X$ and $Y$ are the hydrogen and helium mass fractions, respectively, and the subscript ``$\mathrm{proto\_solar}$'' denotes the protosolar reference values.

\subsection{Parameter Space}
We investigate a multitude of initial parameters with planetary masses between Saturn's mass ($M_{\mathrm{S}}$) and two Jupiter masses ($\MJ$) and various initial specific entropy profiles within 8 and $\SI{11}{\kbperbary}$ (at protosolar composition). To evaluate this plethora of models quantitatively, we investigate the observational accessible atmospheric heavy-element mass fraction $Z_\mathrm{atm}$, which we define as the heavy-element mass fraction of the outermost convective layer, and compare it to the bulk heavy-element mass fraction of the planet $\Zbulk$. Since the atmospheric heavy-element mass fraction is blind toward changes in the core that do not directly influence the envelope, we define a more robust measure, called the heterogeneity $h^2 := \int_{0}^{M} (Z(m) - \Zbulk)^2 \dint{m}$. While $h$ is more robust, it is also more opaque. Hence, we favor $\Zatm/\Zbulk$ in this text. However, we always made sure that both show consistent behavior.
\section{A New Analytic Framework for Mixing}\label{sec:analytic_model}
In this section, we present an analytical prescription that can be used to predict mixing for giant planets with primordial composition gradients. 
Consider a volume element inside a radiative layer at position $r$ that is displaced by a small distance $\Delta r$. Furthermore, let us assume the element moved fast enough such that we can neglect the heat transfer with the surrounding material and assume that its composition remains constant. The density of a volume element $\rho_e$ at $r+\Delta r$ is then given by 
\begin{align}
\rho_e(r + \Delta r) = \rho(r) + \diffp*{\rho}{P}{s,\{X_i\}}\diff{P}{r}\Delta r,
\end{align}
where $P$ is the pressure, $s$ is the specific entropy, and $\{X_i\}$ are the mass fractions of all chemical species. Note that due to the constraint that all mass fractions have to add up to 1, the index $i$ only runs from 1 to $N-1$, where $N$ is the number of chemical species. For the surrounding medium, we must consider the change in entropy and composition. Hence, the change in the surrounding density $\rho_s$ is given by 
\begin{align}
\rho_s(r + \Delta r) = \rho(r) &+ \diffp*{\rho}{P}{s,\{X_i\}}\diff{P}{r}\Delta r\\
&+ \diffp*{\rho}{s}{P,\{X_i\}}\diff{s}{r}\Delta r,\nonumber\\
&+ \sum_{i = 1}^{N-1}\diffp*{\rho}{{X_i}}{P,s,\{X_{j\neq i}\}}\diff{X_i}{r}\Delta r\nonumber.
\end{align}
Let us assume without loss of generality that the volume element is displaced upwards ($\Delta r > 0$). Then, because of buoyancy, the volume element will return to its initial location if $\rho_e(r+\Delta r) > \rho_s(r+\Delta r)$. Since the pressure quickly equalizes between the volume element and the surrounding medium, one can write 
\begin{align}
\diffp*{\rho}{s}{P,\{X_i\}} \diff{s}{r}+ \sum_{i = 1}^{N-1}\diffp*{\rho}{{X_i}}{P,s,\{X_{j\neq i}\}}\diff{X_i}{r} < 0.
\end{align}
Dividing by the first partial derivative, which is negative, and combining the two partial derivatives, we obtain the stability criterion for convection in the presence of composition gradients using the entropy as 
\begin{align}\label{eq:stability_criterion}
\diff{s}{r} - \sum_{i = 1}^{N-1}\diffp*{s}{{X_i}}{P, \rho,\{X_{j\neq i}\}}\diff{X_i}{r} > 0 
\end{align}
\citep[see \textcolor{black}{also}][]{Ledoux_1947, Stevenson_1977b, Bisnovatyi-Kogan_2001, Sur_2024}.
In the absence of composition gradients, Equation \ref{eq:stability_criterion} reduces to the well-known condition \citep[e.g.,][]{Landau_1959}
\begin{align} \label{eq:entropy_Schwarzschild}
\diff{s}{r} > 0.
\end{align}
Since many EoSs are given in terms of $( \rho, T)$, we provide the criterion in terms of constant $T$ instead of $P$ in Appendix \ref{sec:convective_stability_alternative}.

We further simplify the equation by considering only the mass fraction of hydrogen ($X$), helium ($Y$), and all other elements ($Z$). Fixing $Y$ via $Y = 1-X-Z$, Equation \ref{eq:stability_criterion} simplifies to
\begin{align}\label{eq:stability_criterion_mass_XZ}
0 < \diff{s}{m} &- \diffp*{s}{X}{P, \rho, Z}\diff{X}{m}\\
&-\diffp*{s}{Z}{P, \rho, X}\diff{Z}{m}\nonumber.
\end{align}
where we also moved to the (Lagrangian) mass coordinate $m$. While formally equivalent to the Ledoux criterion, Equation \ref{eq:stability_criterion_mass_XZ} allows for a unique insight: An effective entropy gradient stabilizes the system, 
where the composition gradient serves as an additional storage of entropy. Note that the composition also influences the ${ds}/{dm}$ term in Equation \ref{eq:stability_criterion_mass_XZ}. 
The contribution of the composition to the entropy depends on the EoS (i.e., the material properties) of the elements. Typically, heavier chemical species lead to steeper (effective) entropy gradients, and hence are more stable against convection.
For convenience, we define $y_s(t, m)$ as the right-hand side of Equation \ref{eq:stability_criterion_mass_XZ}.
Integrating $y_s(t, m)$ leads to
\begin{align}
    \tilde s(t, m) := s(t, m) + \sXZ(t, m),
\end{align}
where the entropy $s$ is the integral over the entropy derivative (plus an integration constant), and $\sXZ$ is the sum of the integrals over the $X$ and $Z$ derivatives.

Consider now a gas giant at time $t_0$ with an outer convective region ($y_s < 0$) and an inner radiative region ($y_s > 0$), divided by the radiative--convective boundary (RCB) $\mrcb$ ($y_s = 0$). For this planet, $\tilde s$ first rises monotonously until $\mrcb$, after which it falls monotonously. However, the convective envelope is well mixed and the negative entropy gradient is extremely small. Thus, we can approximate the gradients as zero outside the RCB, which allows us to write $\tilde s$ as

\begin{align} \label{eq:s_tilde_approx}
    \tilde s(t_0, m) = \begin{cases}
        s(t_0, m) + \sXZ (t_0, m) & \text{for}~m \leq \mrcb\\
        \senv(t_0) & \text{for}~m > \mrcb,
    \end{cases}
\end{align}
where $\senv(t) = s(t, \mrcb) + \sXZ (t, \mrcb)$.

After time $\dt$, the planet will have cooled down whereby its total entropy decreases by some amount $\dS$. Especially in the early evolution of the planet, cooling is completely dominated by convection. Hence, most of the entropy will be lost in the convective envelope, and we can approximate $\tilde s$ to remain unchanged for $m < \mrcb$. This entropy loss quickly propagates throughout the convective region. As a result, the RCB expands inwards, mixing the composition gradient in the process.
This mixing releases entropy and terminates only once a new RCB with $y_s = 0$ is reached.

\begin{figure}[ht!]
\plotone{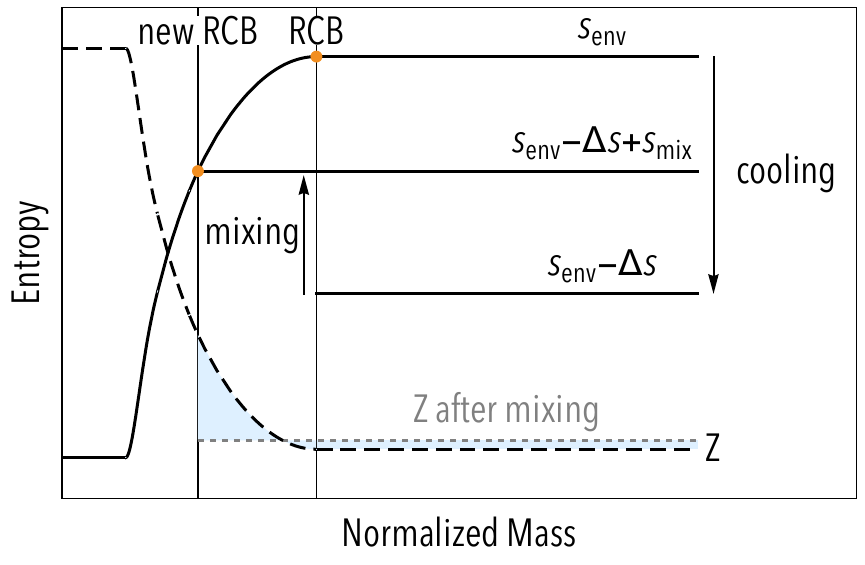}
\caption{A sketch of the idealized mixing process for a positive mixing entropy. The solid black line represents the entropy profiles, and the dashed lines the associated heavy-element mass fraction profile. The light blue region indicates the change in Z that leads to an increase in entropy due to mixing.\label{fig:mixing_visualization}}
\end{figure}

\begin{figure}[ht!]
\plotone{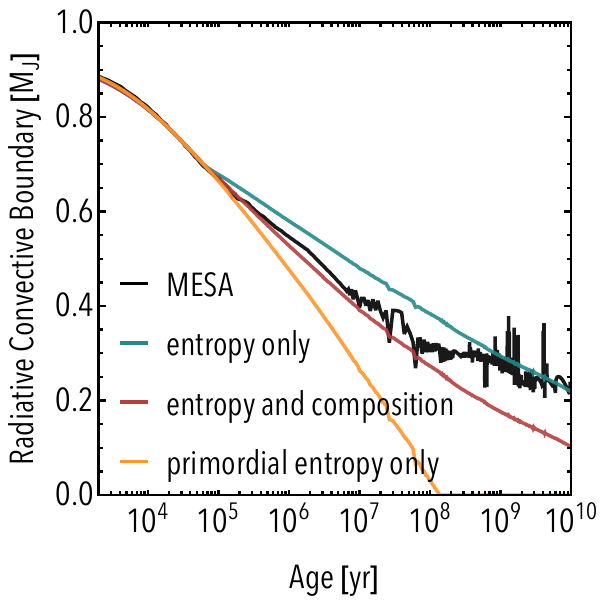}
\caption{The location of the RCB vs. time for a linear primordial entropy gradient from $7 - \SI{10}{\kbperbary}$ and the model ``extended" composition gradient. The black line shows the result using \mesa. The teal line represents our analytical solution assuming $\dsXZ = 0$. The red line is computed in a similar fashion but assuming the initial $\dsXZ$. Lastly, the orange line considers only the primordial entropy, i.e., it ignores any contribution from the composition gradient.\label{fig:analytic_comparison}}
\end{figure}

Since $\tilde s$ in the radiative region is unchanged, the new RCB again fulfills Equation \ref{eq:s_tilde_approx}, i.e., the RCB is always located where the inner effective entropy equals the (decreasing) envelope entropy. A sketch of this process is shown in Fig. \ref{fig:mixing_visualization}. Note, however, that the mixing entropy is typically very small and can be neglected for this derivation (see also Appendix \ref{sec:analytic_mixing_details}).
If we describe this balance in terms of the difference of the surface entropy and the entropy at some mass coordinate $\ds(m) \coloneqq s(M) - s(m)$ (and analogously for $\dsXZ$), we can then write the condition for the RCB location as
\begin{align} \label{eq:RCB_location}
    \frac{\dS}{M - \mrcb} = \ds (\mrcb) + \dsXZ(\mrcb). 
\end{align}
Furthermore, the entropy difference $\ds$ can be interpreted as the sum of the entropy difference from formation (primordial entropy) and the entropy difference from varying the composition (composition entropy). While the sum of the composition entropy and $\dsXZ$ acts stabilizing against convection, the term $\dsXZ$ can be both stabilizing or destabilizing, depending on the sign of the entropy partial derivatives in Equation \ref{eq:stability_criterion_mass_XZ}.
We observed that in our simulations $\dsXZ$ is negative (destabilizing) in the early stages of the evolution and becomes positive (stabilizing) after $\sim \SI{1e8}{\yr}$. 
As limiting cases, we consider the most destabilizing $\dsXZ$ right after formation (i.e., the initial $\dsXZ$) and a vanishing $\dsXZ$ (i.e., $\dsXZ = 0$). 
Figure \ref{fig:analytic_comparison} compares these two approximations to the numerical results obtained using \mesa.
We note that one could also account for the time evolution of $\dsXZ$ to get an estimate that is more consistent with the numerical results.
Overall, we find that our analytic criterion is in good agreement with the numerical simulations. However, the analytic criterion with the initial $\dsXZ$ underestimates the location of the RCB at later stages of the planetary evolution $(\gtrsim \SI{1e8}{\yr})$. This is because $\dsXZ$ becomes positive at these stages. Conversely, the analytic criterion with $\dsXZ = 0$ underestimates the extent of the convective region once the composition gradient starts to be eroded because it ignores the initially destabilizing effect of $\dsXZ$.
Furthermore, Equation \ref{eq:RCB_location} assumes that composition gradients are stabilizing in the absence of entropy gradients. Yet, these so-called semi-convective regions can be overstable, and allow mixing even in the case of a stabilizing composition gradient \citep[e.g.,][see also Section \ref{sec:discussion}]{Kato_1966, Anders_2022, Tulekeyev_2024}.
For example, \citet{Tulekeyev_2024} showed that, depending on the properties of the fluid (i.e., compositional and thermal diffusivity, and kinematic viscosity) and the thermal and compositional stratification of the region, semi-convective regions can either mix completely or remain stratified.
However, these hydrodynamic simulations have not yet reached the scale and conditions required for planetary evolution. Hence, the efficiency of mixing in semi-convective regions under planetary conditions is not yet clear.
Semi-convective regions naturally arise during the planetary evolution in the vicinity of the RCB. Nonetheless, the mixing of these regions can produce Schwarzschild-stable configurations that persist over gigayears. Specifically, these regions become Schwarzschild stable once the post-mixing entropy would be a monotonously increasing function of mass (for details, see Appendix \ref{sec:analytic_mixing_details}). Hence, these regions are still more stable than a system that is solely stabilized by the primordial entropy (see Figure \ref{fig:analytic_comparison}).
As a result, setting $\dsXZ = 0$ gives an upper bound for the RCB, while only considering the primordial entropy gives a lower bound for the RCB. Given that for the planet to become fully mixed the RCB needs to reach the core, i.e., $\mrcb = \SI{0}{\MJ}$, Equation \ref{eq:RCB_location} provides limits for assessing under what conditions a giant planet is fully mixed, within the assumptions of our model (e.g., no rotation; see also Section \ref{sec:discussion_analytic_model}).
\par
Evaluating Equation \ref{eq:RCB_location} in detail requires knowledge of the planetary cooling ($\dS$), the initial primordial entropy profile ($\ds$), and the primordial composition gradient ($\dsXZ$).
While these properties are not well known and are expected to change among different planets, for a rough estimate the criterion is useful. Consider, for example, a composition gradient with \qty{50}{\percent} rock and \qty{50}{\percent} water mixture at $\SI{10}{\kbperbary}$. For this composition, the derivative of the entropy with respect to composition is roughly $\SI{-8}{\kbperbary}$.
Consequently, the composition provides at most a stabilizing effect of roughly $\ds \approx \SI{-8}{\kbperbary} (Z(m) - Z_\mathrm{proto\_solar} $). Given that over evolutionary timescales planets cool to entropies of the order of $\SI{6}{\kbperbary}$, i.e., $\dS/M \approx \SI{4}{\kbperbary}$, we can conclude that any gradient below $Z = 0.5$ will mix over gigayear timescales.
This prediction is consistent with the ``compact'' model in Section \ref{sec:CompGrad}. 
Although our analytical criterion is somewhat simplified, it includes the basic physics required to explore convective mixing in giant planets and is consistent with the numerical simulations.

\section{Results} \label{sec:results}
To explore the behavior of mixing in the planetary interior, we first investigate the influence of the compositional profile, entropy profile, and planetary mass in isolation. Then, we investigate the dependence of the results on the EoS used and the assumed equilibrium temperature.
\subsection{Dependence on the Composition Gradients}\label{sec:CompGrad}
\begin{figure}[ht!]
\plotone{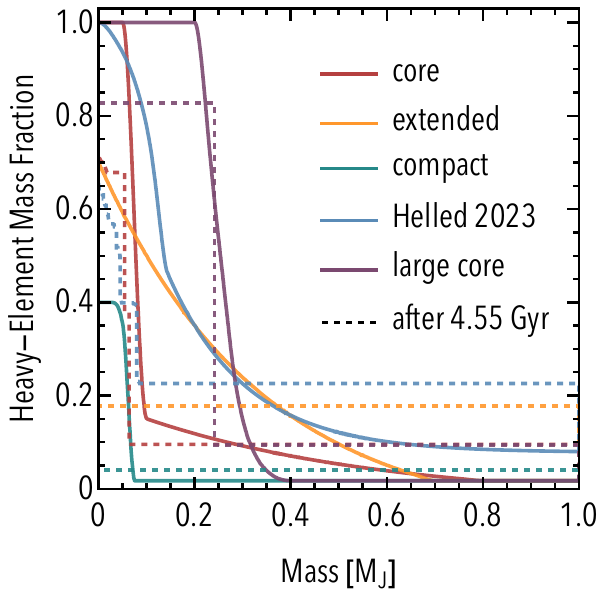}
\caption{The heavy-element distribution for the various models we consider at the beginning (solid lines) and end (dashed lines) of the simulations (see Section \ref{sec:CompGrad} for further details). \label{fig:Z_gradients}}
\end{figure}

We consider a range of composition gradients for a reference model of a $\SI{1}{\MJ}$ planet with a homogeneous entropy of $\SI{10}{\kbperbary}$ before relaxing the heavy-element mass fraction profile. Note that this setup is initially Schwarzschild unstable and thus in the semi-convective regime (see Section \ref{sec:analytic_model}). The initial conditions are not meant to accurately represent a formation scenario, but rather to highlight the different evolution and mixing efficiency for different assumed primordial composition gradients.
The ``core" model represents a high-$Z$ core with a sharp decrease in heavy-element mass fraction, followed by a steady decline toward a constant-envelope heavy-element mass fraction with a protosolar composition. The ``extended'' model begins with $Z = 0.7$ with a continuous decrease in heavy-element mass fraction until protosolar composition is reached. The ``compact'' model has a core of $Z = 0.4$ and a sharp decrease toward protosolar composition. The ``Helled 2023'' model is a metal-rich model with $\Zbulk = 0.25$ motivated by Figure 1 of \citet{Helled2023}. Lastly, the ``large core'' model has a core of $Z = 1$ up to $\SI{0.2}{\MJ} = \SI{64}{\ME}$, followed by a sharp decrease in heavy-element mass fraction. Note that these different primordial composition gradients lead to different planetary radii (see Appendix \ref{sec:radius_evolution} for details).
All the models are evolved for $\SI{10}{Gyr}$. Figure \ref{fig:Z_gradients} shows the heavy-element mass fraction profiles at the initial time and after $\SI{4.55}{Gyr}$ for these simulations.

The numerical experiment shows that the ``compact'' and ``extended" models mix fully, whereas the ``core" and ``Helled 2023" models only mix partially. Moreover, the ``large core" model retains a large fraction of its primordial core throughout the evolution. The results for the ``compact" model suggest that a steep $Z$-gradient alone does not ensure stability against convection. This can be understood using Equation \ref{eq:RCB_location}: If the entropy deficit created by the gradient is not deep enough, even a steep $Z$-gradient eventually gets eroded ($\sim \SI{1e6}{\yr}$ in the case of the ``compact" model).
\par 
Furthermore, the ``core", ``large core", and ``Helled 2023" models have very high $Z$ values and thus very low inner entropy values, retain their cores over evolutionary timescales.
Due to semi-convection being treated as unstable by our numerical experiments, even the cores undergo some mixing. If we would assume that semi-convection is stable, or at least not as efficient as convective layers, this erosion could be reduced significantly (see also Section \ref{sec:analytic_model} and Appendix \ref{sec:analytic_mixing_details}).
We find that convective mixing can lead to the formation of ``convective staircases,": multiple extended convective regions separated by small radiative regions with steep composition gradients.
These staircases evolve in time: some ``steps" disappear and new ones are created. As time progresses, these convective regions can transport heavy elements toward the outer part of the planet and dilute the deep interior with hydrogen and helium (H--He).
\par 
The ``extended" model demonstrates that a heavy-element profile with $Z \sim 0.7$ can still be fully mixed, provided that the $Z$-gradient is not very steep. For these intermediate $Z$ profiles, it is unclear whether all the planetary interior mixes since the efficiency of mixing will depend on the exact conditions. As we discuss in Section \ref{sec:discussion}, this parameter region is also most sensitive to numerical modeling choices (e.g., the used opacity table, the mixing prescription and its parameters). 
\par
The heavy-element mass fraction profile of the ``large core" model remains largely unchanged during the planet's evolution. 
After $\SI{10}{Gyr}$, the atmospheric heavy-element mass fraction is only \qty{35}{\percent} of the bulk heavy-element mass fraction. In other words, the atmospheric abundance is clearly not representative of the bulk composition. However, even for the ``Helled 2023" and ``core" models the atmospheric heavy-element mass fraction is \qty{91}{\percent} and \qty{73}{\percent}, respectively. This highlights the importance of accounting for mixing when interpreting atmospheric abundance measurements. The massive inner core of the ``large core" model would, however, have additional effects like reducing the planetary radius. 

Overall, we can conclude that steep composition gradients with high $Z$ values can remain partially stable over billions of years.
\subsection{Dependence on the Primordial Entropy}\label{sec:primordial_entropy}
\begin{figure}[ht!]
\plotone{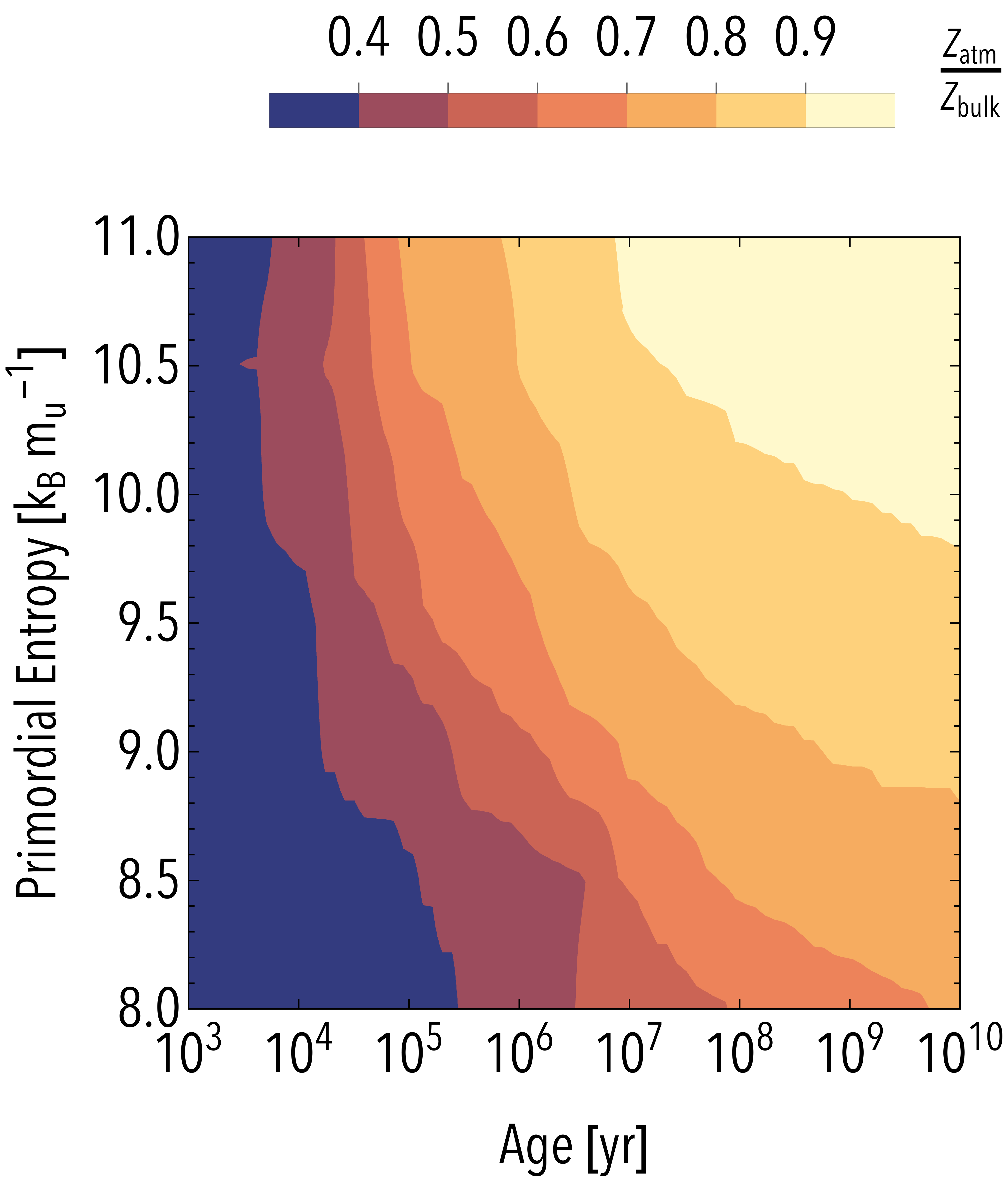}
\caption{The time evolution of $\Zatm/\Zbulk$ for a range of primordial entropies. 
\label{fig:contour_plot_entropy}}
\end{figure}

In this section, we investigate various primordial entropy profiles for a reference model of $\SI{1}{\MJ}$. For all simulations, we use the ``Helled 2023" $Z$ profile to isolate the influence of the primordial entropy.
Again, we focus on studying the impact of the entropy profile on the mixing, and therefore the entropy profiles we consider are not necessarily representative of primordial entropies expected from formation models (see Section \ref{sec:discussion}). 
For simplicity, we use primordial entropy profiles between $\SI{8}{\kbperbary}$ and $\SI{11}{\kbperbary}$ that are constant in mass (prior to relaxing the composition gradient; see Section \ref{sec:numerical_methods}). We chose these entropy values to broadly reflect the expected range from formation models \citep[e.g.,][]{Cumming2018}.
Figure \ref{fig:contour_plot_entropy} shows the evolution of $\Zatm/\Zbulk$. Note that the evolution is shown for timescales ranging between $\num{1e3}$ and $\SI{1e10}{\yr}$, while the ages of most of the observed exoplanets are of the order of gigayears. This means that when we observe exoplanets today, most of the rigorous convective mixing has already taken place. 
Clearly, mixing depends strongly on the primordial entropy. While primordial entropies $\gtrsim \SI{10}{\kbperbary}$ lead to well-mixed interiors with $\Zatm/\Zbulk > 0.9$, lower primordial entropies create more resilient configurations, with $\SI{8}{\kbperbary}$ only leading to $\Zatm/\Zbulk \sim 0.7$. Higher initial entropies lead to shallower entropy profiles, which mix more efficiently (see Section \ref{sec:analytic_model}). Since a higher entropy means faster cooling, a high primordial entropy quickly decreases, which leads to mixing and the erosion of the composition gradient. Crucially, this means that most of the mixing for hot planets occurs early within the first $\sim \SI{1e7}{\yr}$. On the other hand, low primordial entropies lead to deep entropy profiles, which are more stable to convection. Moreover, the planet cools down more slowly, which delays the mixing compared to the high primordial entropies. These cool planets continue to mix over billions of years.

\begin{figure}[ht!]
\plotone{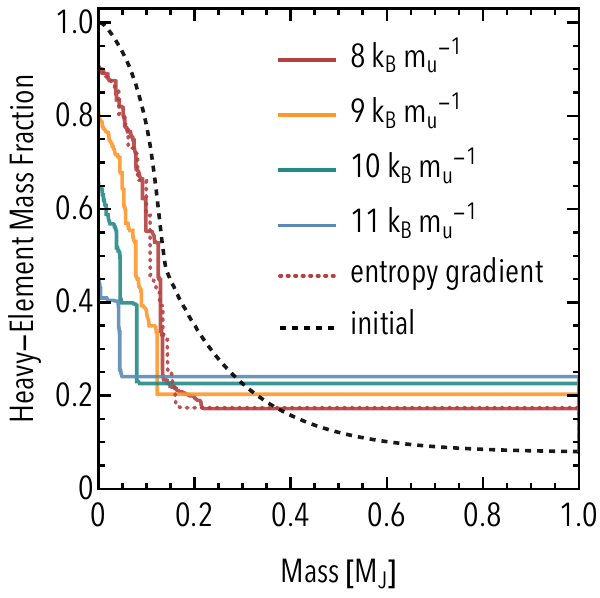}
\caption{The heavy-element distribution for the ``Helled 2023" model at the beginning and end of the simulation for different primordial entropies.\label{fig:initial_entropies}}
\end{figure}

Figure \ref{fig:initial_entropies} presents the heavy-element mass fraction profile for $\SI{8}{\kbperbary}$, $\SI{9}{\kbperbary}$, $\SI{10}{\kbperbary}$, and $\SI{11}{\kbperbary}$ after $\SI{4.55}{Gyr}$. In addition, we investigate an entropy profile that is $\SI{8}{\kbperbary}$ for $m\leq\SI{0.5}{\MJ}$ after which it linearly increases to $\SI{10}{\kbperbary}$ at $m = M$, labeled by ``entropy gradient."
Again, we see that the final heavy-element mass fraction profile is more stable the lower the primordial entropy. Importantly, the ``entropy gradient"  model is virtually unchanged from the homogeneous $\SI{8}{\kbperbary}$ model. While entropy is driving convection, to act (de)stabilizing, the entropy profile must vary in the region of the composition gradient. In the ``entropy gradient" model, the entropy is increased in the outer $m>\SI{0.5}{\MJ}$ of the planet where the composition does not change much. Hence, the only effect of the outer entropy gradient is to delay the onset of mixing. This suggests that in the core-accretion scenario, the convective stability of the innermost part of the planet is unaffected by an increase in entropy of the outer envelope during runaway gas accretion. While such an entropy gradient would delay the onset of mixing, it is not expected to change the outcome over billions of years assuming the accreted gas is of constant composition. 
Therefore, to model convective mixing, and thereby interpret atmospheric metallicities, it is crucial to determine the entropy profile during planetary formation.

\subsection{Dependence on the Planetary Mass}\label{sec:mass_trend}
\begin{figure}[ht!]
\plotone{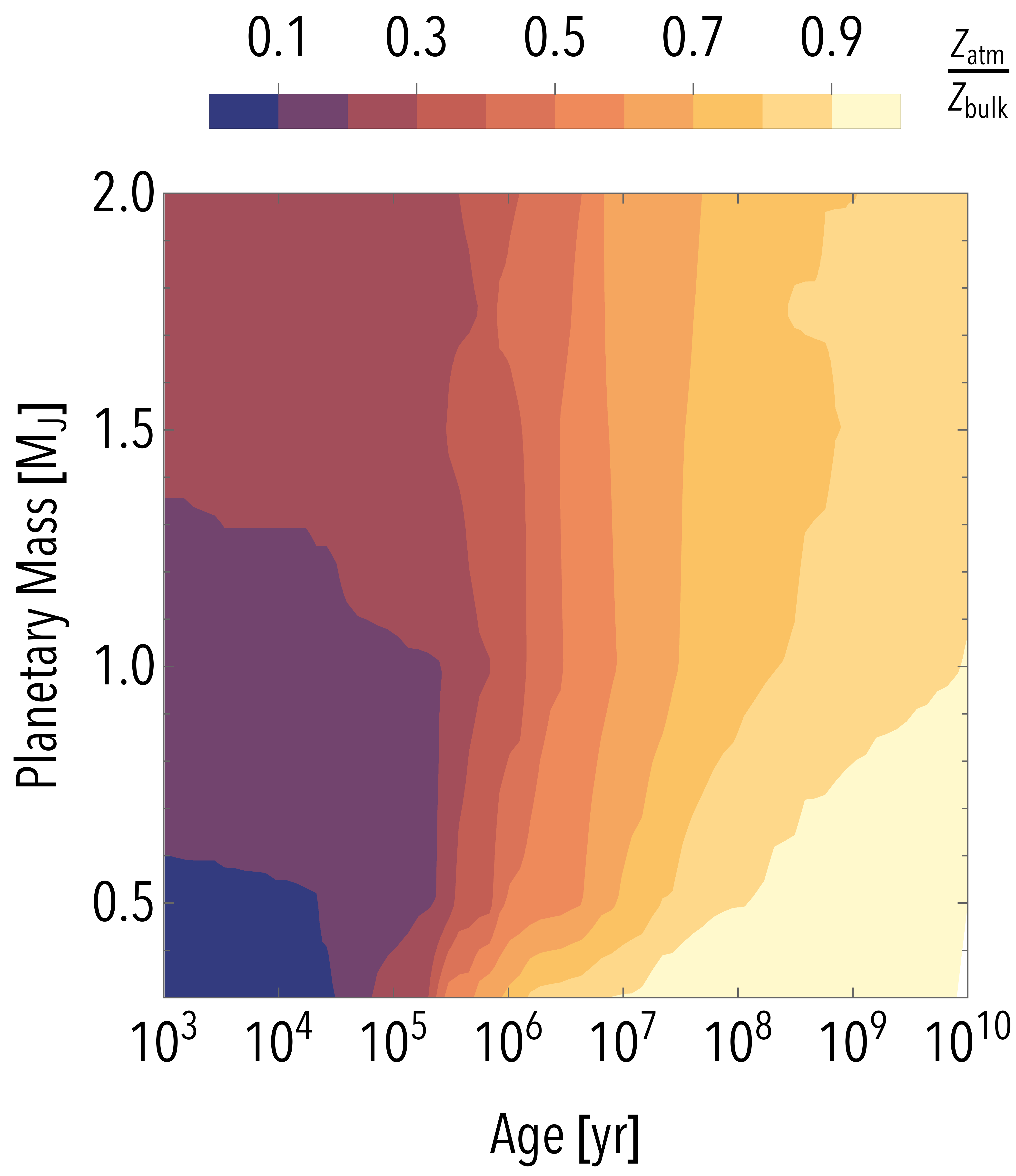}
\caption{The time evolution of $\Zatm/\Zbulk$ for planetary masses between 0.3 and $\SI{2.0}{\MJ}$}. 
\label{fig:contour_plot_mass}
\end{figure}

\begin{figure}[ht!]
\plotone{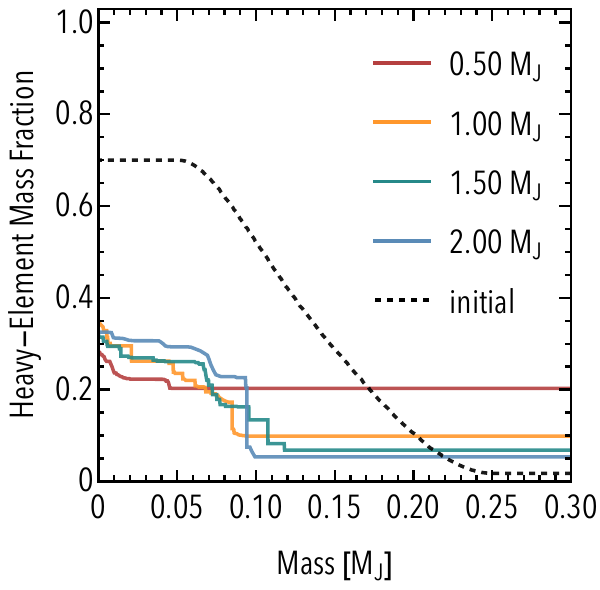}
\caption{The heavy-element mass fraction profiles at the end of the simulations for different planetary masses. The dashed line shows the initial heavy-element mass fraction profile.
\label{fig:Z_profiles_mass}}
\end{figure}

Next, we consider planetary masses between Saturn's mass and two Jupiter masses given the same primordial entropy of $\SI{9}{\kbperbary}$. We employ a heavy-element mass fraction profile with a core of $Z = 0.7$ up to $m = \SI{0.05}{\MJ}$ after which it falls off to protosolar composition at $m = \SI{0.25}{\MJ}$.
This profile, {different from the ``Helled 2023" model in the previous section}, transitions to the envelope composition below $\SI{0.3}{\MJ}$. This allows us to isolate the influence of increasing the envelope mass while keeping the core mass constant. 
It is unknown how the properties of the composition gradient in the planetary deep interior depend on the final planetary mass. From a planet formation perspective, the early stages in giant planet formation (phase 1 and 2) are expected to be similar for different final planetary masses. This is because most of the planetary mass in giant planets is gained during the phase of runaway gas accretion \citep[phase 3; e.g.,][]{Helled2023}. Therefore, it is reasonable to assume the same composition gradients for different final planetary masses. Figure \ref{fig:contour_plot_mass} shows the evolution of $\Zatm / \Zbulk$ and Figure \ref{fig:Z_profiles_mass} highlights the final heavy-element mass fraction profile for $\SI{0.5}{\MJ}$, $\SI{1.0}{\MJ}$, $\SI{1.5}{\MJ}$, and $\SI{2.0}{\MJ}$ after $\SI{4.55}{Gyr}$.

Since we model the envelope with a protosolar composition, and more massive planets have larger envelopes, the inner core will contribute less to the total heavy-element mass for more massive planets. Thus, $\Zatm/\Zbulk$ is initially larger for high-mass planets. However, this changes during the planetary evolution since the lower-mass planets mix faster and more efficiently. The onset of mixing depends on the cooling rate of the planet. Faster cooling leads to lower envelope entropies, which in turn drive the inward growth of the convective region. More massive planets have longer Kelvin-Helmholtz timescales, and thus take longer to mix than their low-mass counterparts. Moreover, because of their larger envelopes, mixing the same core region increases the envelope heavy-element mass fraction less compared to less massive planets. For these less massive planets, the increase in envelope heavy-element mass fraction decreases the envelope entropy noticeably, further destabilizing the inner region and making them more unstable to convection in general (see Section \ref{sec:analytic_model} and Appendix \ref{sec:analytic_mixing_details}).

Combined with the result from Section \ref{sec:primordial_entropy}, this means that if the entropy in the core is the same for planets of different masses, less massive planets would be more mixed. Depending on the formation scenario, however, entropy may increase with planetary mass (see Section \ref{sec:intro}). In this case, the mixing efficiency could increase with planetary mass, producing the opposite trend. The dependence of the primordial composition gradient on the final planetary mass is unknown and is likely to depend on various parameters such as the formation location and accretion rates. This adds a complexity for predicting trends. As a result, a good understanding of the relation between the primordial entropies, the primordial structure, and planetary mass is required. In addition, it would be desirable to consider a larger parameter space of planetary masses with different initial conditions; we hope to address this in future research.

\subsection{Dependence on the Used Equation of State}
\begin{figure}[ht!]
\plotone{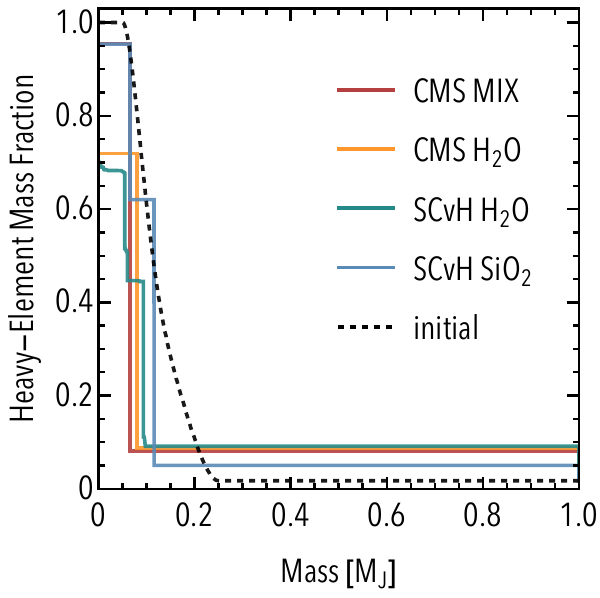}
\caption{The heavy-element mass fraction profiles at \SI{4.55}{Gyr} when using different EoSs for a $\SI{1}{\MJ}$ planet of $\SI{10}{\kbperbary}$.\label{fig:eos}}
\end{figure}

The EoS used for the simulation affects the efficiency of mixing. First of all, the use of different species to represent the heavy elements directly influences the mixing since denser elements typically create larger entropy deficits and are also harder to mix. 
Second, the uncertainty in the H--He EoS and the differences between existing EoSs for these materials also influences the long-term evolution and final internal structure \citep[e.g.,][and references therein]{Helled_2020b}. To demonstrate the effect of the EoS used, we tested the same $Z$ profile for a $\SI{1}{\MJ}$ planet with a homogeneous primordial entropy of $\SI{10}{\kbperbary}$ for four different EoSs: The H--He EoS from \citet[][SCvH]{Saumon1995} with water, SCvH with rock, the H--He EoS of \citet[][CMS]{Chabrier_2021} with water, and CMS with a half-half mixture of water and rock (the default for this study). Figure \ref{fig:eos} compares the final heavy-element mass fraction profiles.
As expected, heavier molecules like SiO$_2$ lead to more stable configurations, which are harder to mix. However, there is also a noticeable discrepancy between the CMS and SCvH EoSs. Namely, models using the CMS EoS mix more efficiently than models using the SCvH EoS. This underlines the importance of using up-to-date EoSs and improves our understanding of the chemical makeup of planetary cores. The fundamental behavior of convective mixing in giant planets investigated here is found to be insensitive to the exact assumed composition and the EoS used. These details, while important for fine-tuning models, do not alter the broader mechanisms of convective mixing, and do not affect the key results presented in this study.

\subsection{Importance of the Atmosphere Model Used and Irradiation}\label{sec:T_eq}
\begin{figure}[ht!]
\plotone{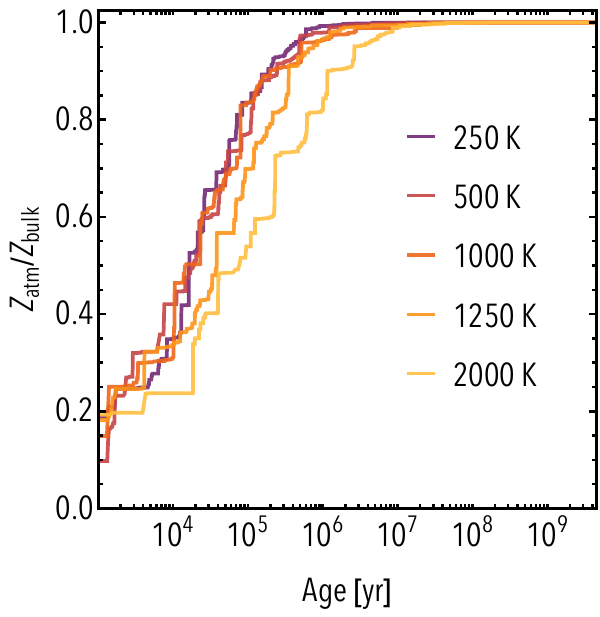}
\caption{The evolution of $\Zatm / \Zbulk$ for different equilibrium temperatures for a $\SI{1}{\MJ}$ planet of $\SI{10}{\kbperbary}$.\label{fig:atm}}
\end{figure}

Planetary atmospheres play a crucial role in the evolution of giant planets \citep[e.g.,][]{Burrows_1997, Fortney_2011, Chen_2023}. In the context of evolution models, they provide the upper boundary conditions and thereby determine the cooling rate of the planet. In \mesa, multiple different atmosphere models are implemented, of which we used the semi-gray atmosphere model by \citet{Guillot2010} with an equilibrium temperature of $T_\mathrm{eq} = \SI{500}{\K}$. This simple model cannot include all the richness of atmospheric physics, especially with respect to cloud formation and chemical modeling more generally. Nevertheless, it captures the essential features of the planetary atmosphere and remains robust across a wide range of parameters, with relative errors in the temperature profile on the order of a few percent \citep[e.g.,][]{Parmentier_2015}.
Figure \ref{fig:atm} shows the evolution of the ``extended" heavy-element mass fraction profile for different equilibrium temperatures using the model from \citet{Guillot2010}. Only very high equilibrium temperatures in excess of $\sim \SI{1250}{\K}$ start to delay the onset of mixing. This simply follows from the slowed-down cooling, which in turn extends the mixing timescale (see Section \ref{sec:mass_trend}). At these high equilibrium temperatures, however, gas giants (i.e., hot Jupiters) are subject to radius inflation \citep{Bodenheimer_2003, Laughlin_2011}. Irrespective of the details of the inflation mechanism, hot Jupiters require special attention that is outside the scope of this study. For all other gas giants, irradiation plays a minor role for the mixing. Thus, while improving \mesa's atmospheric model is a key priority for us in the future, the influence will likely be subdominant compared to other modeling uncertainties like, for example, the EoS.
\section{Discussion} \label{sec:discussion}
While the results shown in this paper appear to be robust, both the numerical and analytical methods are simplified.

\subsection{Numerical Mixing and Stability}
First, convection itself is modeled using mixing-length theory, which is a simplified treatment of convection with free parameters, the most prominent being the mixing length $\alpha$. For the entire study, we fixed $\alpha = 1$. However, larger or smaller values would lead to more/less efficient mixing. Since convective mixing is extremely effective, only very small mixing-length values would significantly alter the outcome of the simulations though.

Second, our simulations are somewhat simplified since the planets are assumed to be isolated and nonrotating. Other processes such as rotation, tides, or magnetohydrodynamic interactions could influence the structure and nature of convection. For example, a recent study by \citet{Fuentes_2023} suggests that rotation can significantly delay the expansion of the convective zone.

Third, \cpm, and our improvement \gm, are both not physically consistent by holding two thermodynamic variables constant throughout the expansion of the convective zone. Although the new modes for pressure and entropy significantly reduce these errors, they still face numerical difficulties with very large heavy-element mass fraction differences. Future studies will have to improve these methods further, to ensure a self-consistent treatment of convective boundaries. In addition, \gm introduces an artificial damping of the mixing for the sake of numerical stability. The reasoning behind \gm is that the Kelvin-Helmholtz timescale is typically much greater than the mixing timescale. Hence, a reduction of the mixing combined with a reduction of the time step should not influence the outcome by too much. We tested the parameters of the algorithm, as well as other simulation parameters like the spatial resolution, extensively to make sure that our simulations are converged.

Fourth, the planet's evolution in nonconvective regions depends on the assumed opacities. The opacity tables we use in this study are \mesa's implementation of the conductive opacities based on \citet{Cassisi2007} and the radiative opacities based on \citet{Freedman2008}.
While these tables cover a large range of temperatures, pressures, and compositions, the atmospheric model is based on a semi-gray approximation. As a result, we cannot model variations in elemental abundances within the atmosphere.
In addition, the Freedman opacity tables do not include the opacity from grains, which may be relevant for cooler atmospheres.
The existence of grains in the atmosphere can delay the planetary cooling \citep{Mueller_2020b, Mueller_2023}.
Future studies could include more sophisticated atmospheric models and investigate how the planetary evolution changes when assuming different atmospheric compositions. 
Nevertheless, despite the importance of the opacity on the evolution, uncertainties in other properties such as the EoS and primordial entropy likely have a greater impact on the evolution and convective mixing. 
\subsection{Analytic Modeling}\label{sec:discussion_analytic_model}
The analytic model developed in Section \ref{sec:analytic_model} is subject to multiple simplifications. The most prominent one is that the model does not treat semi-convection properly. Therefore, regions that are Schwarzschild unstable, but not Ledoux unstable, might (partially) mix in reality while being treated as stable by our analytical model. This can also be observed in, for example, the ``Helled 2023" model in Figure \ref{fig:Z_gradients}, where the inner core is partially eroded into the envelope but would be stable according to Equation \ref{eq:RCB_location}.
In line with the numerical methods, we did not consider any additional effects like rotation that could affect the evolution of the convective regions.
Furthermore, we assume that the outer envelope remains convective throughout the planet's evolution. However, depending on the opacity of the material, radiative regions might form during the late stages of the evolution. We hope to investigate all these effects in future studies.

\subsection{Connection to Planet Formation} \label{sec:connection_to_reality}
In this study, we isolated the dependence of convective mixing on various planetary parameters without favoring any particular formation mechanism (see Section \ref{sec:intro}). We showed that the mixing outcome depends strongly on the primordial entropy and composition gradient. In reality, both the composition and entropy are shaped by the formation mechanism. As a consequence, the formation mechanism determines the mixing inside the planet, which in turn determines the atmospheric heavy-element mass fraction. This means that different formation pathways may lead to different atmospheric heavy-element mass fractions. Moreover, linking atmospheric abundances to the planetary origin requires a more comprehensive modeling approach that properly links the planetary formation and evolution.

\subsection{Connection to Observations}
This study primarily explored the general mechanisms driving convective mixing in giant planets, emphasizing the fundamental theoretical properties. Future research should assess how the results can be directly linked to observational data.
In addition, we showed that most of the mixing occurs within the first $\sim$ $\SI{1e7}{\yr}$. 
We note that most observed exoplanets have ages of several $\SI{1e9}{\yr}$, with the exception of planets detected by direct imagining, which could have ages as young as $\SI{1e7}{\yr}$. 
This makes the early ``mixing" epoch of hot planets observationally inaccessible.
Furthermore, the atmospheric heavy-element mass fraction $\Zatm$ referenced throughout this study corresponds to the outer convective envelope. Depending on the exact physical conditions, the outer RCB is typically located somewhere between $1$ and $\SI{1000}{\bar}$ in our simulations. Observations, however, often only probe the millibar to bar regime.
Moreover, the heavy elements in this work are represented by either pure rock, pure water, or a mixture of the two, while atmospheric measurements provide information regarding the abundances of specific molecules that are used as proxies for the overall atmospheric metallicity \citep[for a review, see][]{Madhusudhan_2019}.
Therefore, although we infer well-defined trends for the atmospheric heavy-element mass fraction, interpreting these results observationally should be approached with caution and should be investigated in detail in future research. 

\section{Summary and Conclusions} \label{sec:conclusions}
\begin{figure*}[ht!]
\plotone{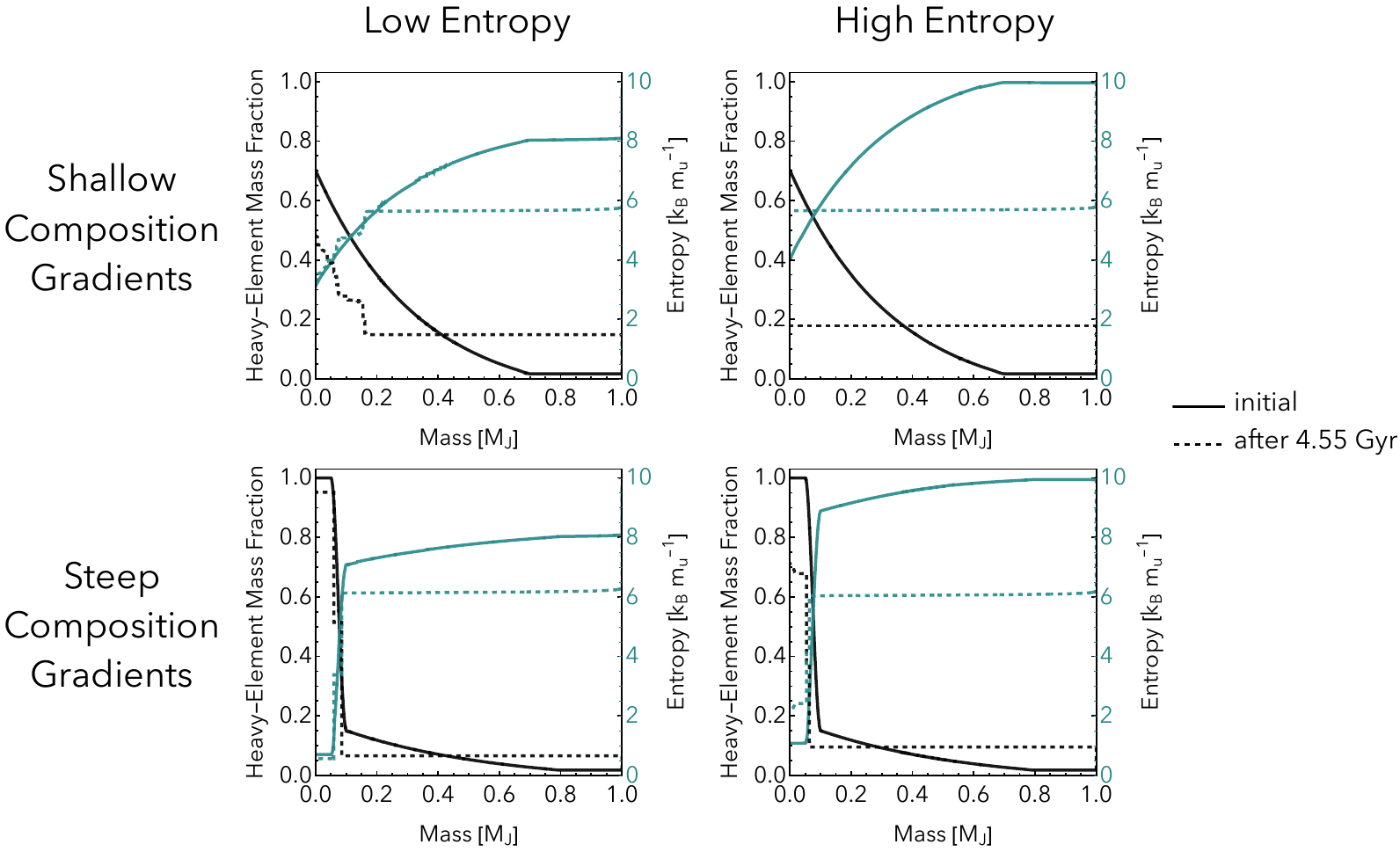}
\caption{Final heavy-element mass fraction profiles for four different initial setups sorted by their convective stability. The figures in the low-entropy column started with a primordial entropy of $\SI{8}{\kbperbary}$ and the figures in the high-entropy column with $\SI{10}{\kbperbary}$ (for details of the entropy profiles, see Section \ref{sec:initial_model}).\label{fig:overview}}
\end{figure*}
In this study, we developed new numerical and analytical tools for modeling convective mixing in gas giant planets. Furthermore, we investigated the connection between the atmospheric composition and the bulk planetary heavy-element mass fraction.
We simulated the evolution of primordial composition and entropy gradients over $\SI{10}{Gyr}$ using updated EoSs for water, rock, and hydrogen-helium. We investigated planetary masses between the mass of Saturn and two Jupiter masses, as well as various primordial entropy and heavy-element mass fraction profiles. Our study shows that convective mixing, and therefore also the final atmospheric composition, are controlled by the planetary primordial entropy. Figure \ref{fig:overview} summarizes our main findings. We find that low primordial entropies $(\lesssim \SI{9}{\kbperbary})$ and deep heavy-element mass fraction profiles $(Z\gtrsim 0.8)$ create stable configurations over evolutionary timescales. On the other hand, high primordial entropies $(\gtrsim \SI{10}{\kbperbary})$ and shallow heavy-element mass fraction profiles $(Z\lesssim 0.6)$ lead to a fully mixed planet. In intermediate cases, the details matter, and both the high-entropy deep-$Z$ profiles and the low-entropy shallow-$Z$ profiles can be partially eroded over evolutionary timescales.

This highlights the importance of understanding the primordial structure of gas giants. Although the mixing of each individual planet is a complicated, multivaried problem, the overall trends highlighted above, as well as the analytic framework developed, appear to be robust.

The key conclusions from this study can be summarized as follows:
\begin{itemize}
    \item Low internal entropies $(\lesssim \SI{9}{\kbperbary})$ and large compositional differences in the planetary deep interior $(Z\gtrsim 0.8)$ lead to nonuniform compositions. This can drastically influence the ratio of atmospheric to bulk heavy-element mass fraction.
    
    \item Giant planets with high primordial internal entropy $(\gtrsim \SI{10}{\kbperbary})$ and extended, shallow composition gradients $(Z\lesssim 0.6)$ will mix rapidly $(\sim \SI{1e7}{\yr})$.
    
    \item The primordial entropy profile dictates the mixing in giant planets. Higher primordial entropies lead to faster cooling and more efficient convective mixing. Therefore, hot planets typically mix within $\sim\SI{1e7}{\yr}$ while in colder planets mixing can take place over gigayears. A hot (high-entropy) envelope does not guarantee rapid mixing since the deep interior can have significantly lower entropy.
    
    \item The scaling of mixing efficiency with mass is governed by the primordial entropy. For the same primordial entropy and core mass, low-mass planets mix more efficiently than high-mass planets. If the primordial core entropy would increase with mass, however, high mass-planets may mix more efficiently. Understanding this trend is crucial for revealing the relationship between planetary mass and mixing efficiency.
    
     \item Constraining the primordial entropy from simulations and observations is critical for determining the evolution and internal structures of giant planets.
    
\end{itemize}

Our study is a crucial step toward interpreting the atmospheric measurements of giant exoplanets, and thereby can help to uncover the mystery of their origin. Further theoretical investigations such as advanced hydrodynamical simulations of convective mixing with composition gradients and planet formation simulations are important for understanding the efficiency of mixing inside giant planets.
\par 

In addition, this study clearly demonstrates that the primordial entropy profile, which is determined by the formation history of the planet, controls the planetary evolution and internal structure. As a result, we now have a new link between internal structure and origin. 
The upcoming atmospheric measurements from JWST, ARIEL, and ground-based observations can therefore unveil new information on giant planet formation. 

\begin{acknowledgments}
We thank Simon Müller for valuable discussions. This work has been carried out within the framework of the National Centre of Competence in Research PlanetS supported by the Swiss National Science Foundation under grant Nos. \texttt{51NF40\_182901} and \texttt{51NF40\_205606}.
\end{acknowledgments}

\software{\mesa \citep{Paxton_2011, Paxton_2013, Paxton_2015, Paxton_2018, Paxton_2019, Jermyn_2023},
\texttt{py\_mesa\_reader} (\url{https://github.com/wmwolf/py_mesa_reader}),
\texttt{NumPy} \citep{harris2020array},
\texttt{SciPy} \citep{2020SciPy-NMeth}
}

\appendix
\section{Improvements to \mesa}\label{sec:improvements_to_mesa}
As mentioned in the main text, for this work we extended \mesa's \cpm ~algorithm. 
Inside a gas giant planet, the energy transport is either dominated by convection or by radiation and conduction. In layers that are unstable to convection (convective layers in short), transport of energy and material occurs very rapidly, with mixing timescales of $t_\mathrm{mix,conv} \sim 10$--$\SI{100}{\yr}$. In contrast, in layers that are stable against convection (e.g., radiative/conductive layers) energy and material transport is very slow ($t_\mathrm{mix,diff} \sim \num{1e11}$--$\SI{1e12}{\yr}$). Hence, determining the boundary between convective and radiative regions is crucial for modeling mixing accurately. \mesa's \cpm algorithm \citep{Paxton_2019} already improves upon the commonly employed ``sign-change" algorithm where an RCB is determined by a change of sign in the difference between radiative and Ledoux gradient $y = \nabla_\mathrm{rad} - \nabla_\mathrm{L}$ between two neighboring cells. The sign-change algorithm typically underestimates convective layer sizes and can lead to physical inconsistencies, especially when encountering composition discontinuities \citep{Gabriel_2014}. The \cpm algorithm remedies many of the shortcomings of the sign-change approach by expanding convective regions until they reach their full extent. During this additional (predictive) mixing, however, the code holds either temperature and pressure or temperature and density constant. For planets, the temperature can change drastically with changes in composition. Therefore, holding the temperature constant while changing the composition often leads to convergence issues of \mesa's solver. Hence, we introduced the two new modes described in Section \ref{sec:numerical_methods} to improve the convergence.

As described in the main text, \gm monitors the change in composition of the planet. More specifically, we evaluate the mean squared distance: 
\begin{align} \label{eq:Z_MSD}
\sigma_\mathrm{sol}^2 = \int_{0}^{M} (Z(t+\dt, m) - Z(t, m))^2 \frac{dm}{M},
\end{align}
where $Z$ is the heavy-element mass fraction, $M$ is the mass of the planet, $m$ is the mass coordinate, $t$ the time coordinate, and $\dt$ the time step.
If $\sigma_\mathrm{sol}$ is larger than a user-supplied critical value $\sigma_\mathrm{sol,crit}$, the mixing efficiency is reduced and the solver redoes the calculation. This process is repeated until $\sigma_\mathrm{sol}<\sigma_\mathrm{sol,crit}$.

Similarly, \gm prevents the expansion of the convective boundary during \cpm beyond $\sigma_\mathrm{cpm,crit}$, i.e., $\sigma_\mathrm{cpm} < \sigma_\mathrm{cpm,crit}$, where $\sigma_\mathrm{cpm}$ is computed analogously to Equation \ref{eq:Z_MSD} comparing the pre- and post-\cpm heavy-element mass fraction profiles.

Furthermore, \gm provides a mode that monitors the maximum abundance change in \mesa's cells in addition to the mean squared distance of the profiles. If mixing was inhibited by any of the mentioned boundaries, the time step is reduced. Effectively, \gm mixes the same region in a few smaller time steps rather than all of it in one large time step. 

Besides these functionalities, \gm comes with some additional features and options. In contrast to \texttt{mesa\_custom\_eos}, \gm modifies \mesa's source code directly. While we strive to make \gm open source in the future, it is currently still under development. Thus, all options highlighted below are still subject to change.

First, the user can pick which isotope should be monitored by \gm, where every isotope used in the reaction network is valid. Next, many additional checks can be performed to ensure the stability of a simulation. Ideally, \gm checks the resulting model after a time step and repeats the calculation with a damped mixing diffusion coefficient, if required. However, sometimes \mesa's solver crashes before finishing even one tentative step because the model simply mixes too much. In this case, \mesa repeats the calculation with a smaller time step (called a retry). Since convective mixing is extremely efficient in giant planets, however, even very small time steps might not improve convergence. Thus, if the time step becomes too small, \gm reduces the mixing efficiency before taking a step. This reduction increases with decreasing time step until mixing is completely turned off (including \cpm). This option often helps the solver to recover and to mix very challenging periods in the planet's evolution. Another issue can arise when \cpm tries to expand a convective region into the radiative atmosphere. Often, the expanding convective layer can cause a negative surface luminosity, which leads the solver to do a retry. However, the mass fraction in the atmosphere is completely negligible for the planetary bulk composition. Thus, \gm allows the user to define an ``exclusion zone" where no expansion of the convective region during \cpm is allowed. For example, one could exclude the outer \qty{1}{\percent} of the planetary mass. In some cases, this drastically decreases the simulation runtime.

\subsection{\mh}
To simplify the setup, the monitoring, and the evaluation of \mesa simulations, we developed a Python package called \mh. The package allows the user, among other things, to modify namelist files, create initial entropy and compositional gradient profiles readable by \mesa, and provides a multitude of additional analyzing functionalities, all using Python.

Namelists can be saved, modified, and restored to their original status. For example, the package supports Python's \texttt{with} block, where a namelist is modified within the code block but automatically restored after the block is exited. It also provides methods that automatically name output folders according to user-specified parameters. These features are especially handy when dealing with a large number of simulations. In addition, the user can set values like the initial mass in units of Jupiter mass rather than grams or solar masses or the initial entropy in units of $\si{\kbperbary}$. The package also takes care of converting a Python function for the initial heavy-element mass fraction or the initial entropy into a \mesa-readable file. The code comes also with a number of predefined composition profiles, including the ones used in this study. Building on \texttt{py\_mesa\_reader}, the package comes with a multitude of tools to analyze a large number of simulations simultaneously.\footnote{\url{https://github.com/wmwolf/py_mesa_reader}} The user could, for example, plot the final heavy-element mass fraction profile of a set of simulations with a single line of code or extract the atmospheric heavy-element mass fraction after $\SI{1}{Gyr}$ for all simulations and export it as a CSV file. The full documentation, as well as the code itself, can be found on GitHub\footnote{\url{https://github.com/Henrik-Knierim/py_mesa_helper}}\citep{py_mesa_helper}.
\section{Alternative Formulation of Convective Stability}\label{sec:convective_stability_alternative}
For $ s = s(P, \rho, \{X_i\})$, the total differential is given by
\begin{align}
\dint s = \diffp*{s}{P}{ \rho,\{X_i\}}\dint P
+ \diffp*{s}{ \rho}{P,\{X_i\}}\dint \rho
+ \sum_{i = 1}^{N-1}\diffp*{s}{{X_i}}{P, \rho,\{X_{j\neq i}\}}\dint X_i.
\end{align}
Interpreting $s$ as a function of temperature instead of pressure, we obtain
\begin{align}
\dint s = \diffp*{s}{T}{ \rho,\{X_i\}}\dint T
+ \diffp*{s}{ \rho}{T,\{X_i\}}\dint \rho
+ \sum_{i = 1}^{N-1}\diffp*{s}{{X_i}}{T, \rho,\{X_{j\neq i}\}}\dint X_i,
\end{align}
where the total differential of the temperature is given by
\begin{align}
\dint T = \diffp*{T}{P}{ \rho,\{X_i\}}\dint P
+ \diffp*{T}{ \rho}{P,\{X_i\}}\dint \rho
+ \sum_{i = 1}^{N-1}\diffp*{T}{{X_i}}{P, \rho,\{X_{j\neq i}\}}\dint X_i.
\end{align}
Solving the system for the partial derivative in Equation \ref{eq:stability_criterion} yields
\begin{align}
\diffp*{s}{{X_i}}{P, \rho,\{X_{j\neq i}\}} = \diffp*{s}{{X_i}}{T, \rho,\{X_{j\neq i}\}} + 
\diffp*{s}{T}{ \rho,\{X_i\}} \diffp*{T}{{X_i}}{P, \rho,\{X_{j\neq i}\}}
\end{align}
Now, we can use the properties of the Jacobi to arrive at 
\begin{align}
\diffp*{s}{{X_i}}{P, \rho,\{X_{j\neq i}\}} &= \diffp*{s}{{X_i}}{T, \rho,\{X_{j\neq i}\}}
- \frac{\diffp*{s}{T}{\rho,\{X_{i}\}}}{\diffp*{P}{T}{\rho,\{X_{i}\}}}
\diffp*{P}{{X_i}}{T, \rho,\{X_{j\neq i}\}}.
\end{align}
This relation can be substituted into Equation \ref{eq:stability_criterion} to obtain the stability criterion in terms of constant temperature and density instead of pressure and density.
\section{Further Details of the Analytic Mixing Model}\label{sec:analytic_mixing_details} 
The mixing entropy we neglect in Section \ref{sec:analytic_model} can be approximated by 
\begin{align}
    s_\mathrm{mix} = \frac{1}{M - \mrcb + \delta m}\int_{\mrcb - \delta m}^{M} \int_{Z(m)}^{Z_\mathrm{env}}\diff{s}{Z}\dint{Z}\dint{m},
\end{align}
where $\mrcb$ is the RCB before expanding the boundary, $\delta m$ is the mass by which we expand the boundary, and $Z_\mathrm{env}$ is the mean value of the heavy-element mass fraction in this region. Importantly, since we compute the mean value over the expanded region, the entropy gained in the $Z(m)>Z_\mathrm{env}$ region partially cancels with the entropy lost in the $Z(m)<Z_\mathrm{env}$ region. Thus, $s_\mathrm{mix}$ is typically of the order of $|s_\mathrm{mix}| \sim \SI{1e-4}{\kbperbary}$, where the sign of $s_\mathrm{mix}$ depends on the exact shape of $\diff{s}{Z}$. For the expanding convective region, this term is negligible in comparison to the other entropy terms. However, the mixing entropy plays a crucial role in semi-convective regions. 
Assuming that semi-convection is unstable (as \mesa's \cpm does), the only potentially stabilizing term is the mixing entropy.

\begin{figure}[ht!]
\plotone{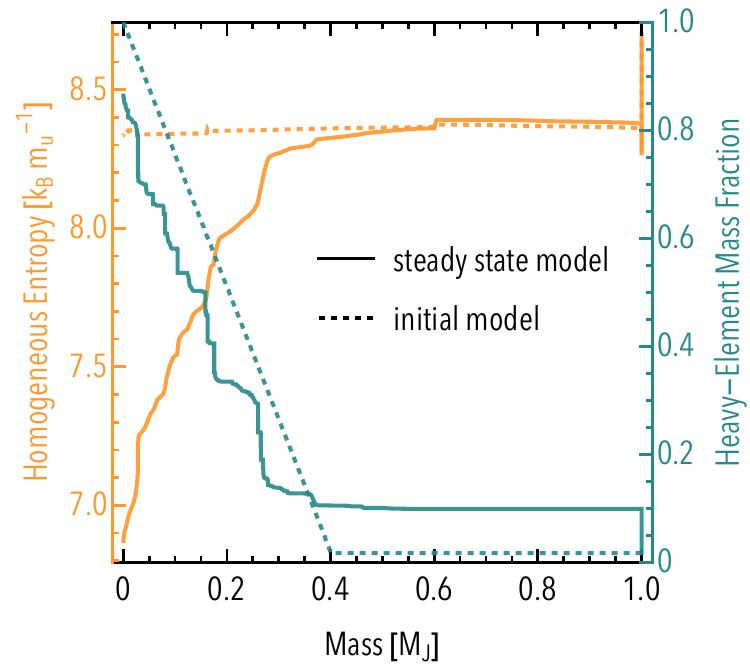}
\caption{Entropy assuming a homogeneous composition and the associated heavy-element mass fraction for the model from Appendix \ref{sec:analytic_mixing_details} at the beginning and after reaching steady state.\label{fig:entropy_comparison}}
\end{figure}

Moreover, the planet will be stable against convection in a mass interval $[m_1, m_2]$ if the post-mixing entropy profile is monotonously increasing with mass.
Figure \ref{fig:entropy_comparison} shows the post-mixing entropy for a planet with a linear composition profile from $Z(m = \SI{0.0}{\MJ}) = 1.0$ to $Z(m =\SI{0.4}{\MJ}) = Z_\mathrm{proto\_solar}$.
For this model, we fixed the entropy in the envelope, allowing the system to reach a steady state. In this case, the entropy of a fully mixed model increases monotonously with mass.
\section{Radius Evolution}\label{sec:radius_evolution}
\begin{figure}[ht!]
\plotone{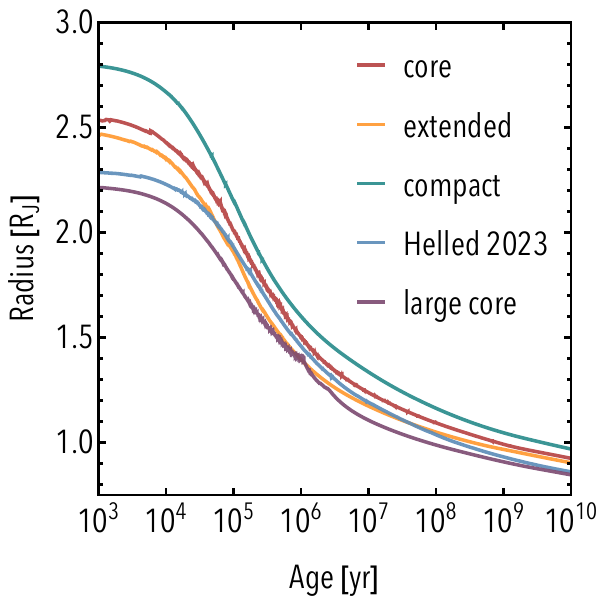}
\caption{Radius evolution of the models presented in Section \ref{sec:CompGrad}\label{fig:radius_evolution}.}
\end{figure}

In Section \ref{sec:CompGrad}, we investigated the mixing of different primordial composition gradients for the same planetary mass ($\SI{1}{\MJ}$) and entropy ($\SI{10}{\kbperbary}$). Even at the same entropy, however, different primordial composition gradients lead to different interior structures, and therefore different radii. Figure \ref{fig:radius_evolution} shows the radius evolution of the models investigated in Section \ref{sec:CompGrad}. 
The different models begin with different initial planetary radii where the contraction depends on the mixing and cooling of the planet. The figure clearly shows that the primordial structure affects the planetary radius, especially at young ages. As a result, measuring the planetary radius of young giant planets can reveal important information on their formation path and primordial thermal state. We hope to investigate this topic in detail in a future study.
\bibliography{main}{}

\begin{thebibliography}{}
\expandafter\ifx\csname natexlab\endcsname\relax\def\natexlab#1{#1}\fi
\providecommand{\url}[1]{\href{#1}{#1}}
\providecommand{\dodoi}[1]{doi:~\href{http://doi.org/#1}{\nolinkurl{#1}}}
\providecommand{\doeprint}[1]{\href{http://ascl.net/#1}{\nolinkurl{http://ascl.net/#1}}}
\providecommand{\doarXiv}[1]{\href{https://arxiv.org/abs/#1}{\nolinkurl{https://arxiv.org/abs/#1}}}

\bibitem[{{Anders} {et~al.}(2022){Anders}, {Jermyn}, {Lecoanet}, {Fraser},
  {Cresswell}, {Joyce}, \& {Fuentes}}]{Anders_2022}
{Anders}, E.~H., {Jermyn}, A.~S., {Lecoanet}, D., {et~al.} 2022, \apjl, 928,
  L10, \dodoi{10.3847/2041-8213/ac5cb5}

\bibitem[{{Aurnou} {et~al.}(2020){Aurnou}, {Horn}, \& {Julien}}]{Aurnou_2020}
{Aurnou}, J.~M., {Horn}, S., \& {Julien}, K. 2020, Physical Review Research, 2,
  043115, \dodoi{10.1103/PhysRevResearch.2.043115}

\bibitem[{{Berardo} \& {Cumming}(2017)}]{Berardo2017b}
{Berardo}, D., \& {Cumming}, A. 2017, \apjl, 846, L17,
  \dodoi{10.3847/2041-8213/aa81c0}

\bibitem[{{Berardo} {et~al.}(2017){Berardo}, {Cumming}, \&
  {Marleau}}]{Berardo2017a}
{Berardo}, D., {Cumming}, A., \& {Marleau}, G.-D. 2017, \apj, 834, 149,
  \dodoi{10.3847/1538-4357/834/2/149}

\bibitem[{{Bisnovatyi-Kogan}(2001)}]{Bisnovatyi-Kogan_2001}
{Bisnovatyi-Kogan}, G.~S. 2001, {Stellar physics. Vol.1: Fundamental concepts
  and stellar equilibrium} ((Berlin: Springer))

\bibitem[{{Bodenheimer} {et~al.}(2003){Bodenheimer}, {Laughlin}, \&
  {Lin}}]{Bodenheimer_2003}
{Bodenheimer}, P., {Laughlin}, G., \& {Lin}, D. N.~C. 2003, \apj, 592, 555,
  \dodoi{10.1086/375565}

\bibitem[{{Bodenheimer} {et~al.}(2018){Bodenheimer}, {Stevenson}, {Lissauer},
  \& {D'Angelo}}]{Bodenheimer_2018}
{Bodenheimer}, P., {Stevenson}, D.~J., {Lissauer}, J.~J., \& {D'Angelo}, G.
  2018, \apj, 868, 138, \dodoi{10.3847/1538-4357/aae928}

\bibitem[{Burrows {et~al.}(1997)Burrows, Marley, Hubbard, Lunine, Guillot,
  Saumon, Freedman, Sudarsky, \& Sharp}]{Burrows_1997}
Burrows, A., Marley, M., Hubbard, W.~B., {et~al.} 1997, The Astrophysical
  Journal, 491, 856, \dodoi{10.1086/305002}

\bibitem[{{Cassisi} {et~al.}(2007){Cassisi}, {Potekhin}, {Pietrinferni},
  {Catelan}, \& {Salaris}}]{Cassisi2007}
{Cassisi}, S., {Potekhin}, A.~Y., {Pietrinferni}, A., {Catelan}, M., \&
  {Salaris}, M. 2007, \apj, 661, 1094, \dodoi{10.1086/516819}

\bibitem[{{Chabrier} \& {Debras}(2021)}]{Chabrier_2021}
{Chabrier}, G., \& {Debras}, F. 2021, \apj, 917, 4,
  \dodoi{10.3847/1538-4357/abfc48}

\bibitem[{{Chen} {et~al.}(2023){Chen}, {Burrows}, {Sur}, \&
  {Arevalo}}]{Chen_2023}
{Chen}, Y.-X., {Burrows}, A., {Sur}, A., \& {Arevalo}, R.~T. 2023, \apj, 957,
  36, \dodoi{10.3847/1538-4357/acf456}

\bibitem[{{Cumming} {et~al.}(2018){Cumming}, {Helled}, \&
  {Venturini}}]{Cumming2018}
{Cumming}, A., {Helled}, R., \& {Venturini}, J. 2018, \mnras, 477, 4817,
  \dodoi{10.1093/mnras/sty1000}

\bibitem[{{Dewberry}(2023)}]{Dewberry_2023}
{Dewberry}, J.~W. 2023, \mnras, 521, 5991, \dodoi{10.1093/mnras/stad546}

\bibitem[{{Fortney} {et~al.}(2011){Fortney}, {Ikoma}, {Nettelmann}, {Guillot},
  \& {Marley}}]{Fortney_2011}
{Fortney}, J.~J., {Ikoma}, M., {Nettelmann}, N., {Guillot}, T., \& {Marley},
  M.~S. 2011, \apj, 729, 32, \dodoi{10.1088/0004-637X/729/1/32}

\bibitem[{{Freedman} {et~al.}(2008){Freedman}, {Marley}, \&
  {Lodders}}]{Freedman2008}
{Freedman}, R.~S., {Marley}, M.~S., \& {Lodders}, K. 2008, \apjs, 174, 504,
  \dodoi{10.1086/521793}

\bibitem[{Fuentes {et~al.}(2023)Fuentes, Anders, Cumming, \&
  Hindman}]{Fuentes_2023}
Fuentes, J.~R., Anders, E.~H., Cumming, A., \& Hindman, B.~W. 2023, The
  Astrophysical Journal Letters, 950, L4, \dodoi{10.3847/2041-8213/acd774}

\bibitem[{{Fuentes} \& {Cumming}(2020)}]{Fuentes_2020}
{Fuentes}, J.~R., \& {Cumming}, A. 2020, Physical Review Fluids, 5, 124501,
  \dodoi{10.1103/PhysRevFluids.5.124501}

\bibitem[{{Fuentes} {et~al.}(2022){Fuentes}, {Cumming}, \&
  {Anders}}]{Fuentes_2022}
{Fuentes}, J.~R., {Cumming}, A., \& {Anders}, E.~H. 2022, Physical Review
  Fluids, 7, 124501, \dodoi{10.1103/PhysRevFluids.7.124501}

\bibitem[{{Fuller}(2014)}]{Fuller_2014}
{Fuller}, J. 2014, \icarus, 242, 283, \dodoi{10.1016/j.icarus.2014.08.006}

\bibitem[{Gabriel {et~al.}(2014)Gabriel, Noels, Montalb{\'a}n, \&
  Miglio}]{Gabriel_2014}
Gabriel, M., Noels, A., Montalb{\'a}n, J., \& Miglio, A. 2014, Astronomy {\&}
  Astrophysics, 569, A63, \dodoi{10.1051/0004-6361/201423442}

\bibitem[{{Gardner} {et~al.}(2006){Gardner}, {Mather}, {Clampin}, {Doyon},
  {Greenhouse}, {Hammel}, {Hutchings}, {Jakobsen}, {Lilly}, {Long}, {Lunine},
  {McCaughrean}, {Mountain}, {Nella}, {Rieke}, {Rieke}, {Rix}, {Smith},
  {Sonneborn}, {Stiavelli}, {Stockman}, {Windhorst}, \& {Wright}}]{Gardner2006}
{Gardner}, J.~P., {Mather}, J.~C., {Clampin}, M., {et~al.} 2006, \ssr, 123,
  485, \dodoi{10.1007/s11214-006-8315-7}

\bibitem[{{Guillot}(2010)}]{Guillot2010}
{Guillot}, T. 2010, \aap, 520, A27, \dodoi{10.1051/0004-6361/200913396}

\bibitem[{{Hands} \& {Helled}(2022)}]{Hands_2022}
{Hands}, T.~O., \& {Helled}, R. 2022, \mnras, 509, 894,
  \dodoi{10.1093/mnras/stab2967}

\bibitem[{Harris {et~al.}(2020)Harris, Millman, van~der Walt, Gommers,
  Virtanen, Cournapeau, Wieser, Taylor, Berg, Smith, Kern, Picus, Hoyer, van
  Kerkwijk, Brett, Haldane, del R{'{\i}}o, Wiebe, Peterson,
  G{'{e}}rard-Marchant, Sheppard, Reddy, Weckesser, Abbasi, Gohlke, \&
  Oliphant}]{harris2020array}
Harris, C.~R., Millman, K.~J., van~der Walt, S.~J., {et~al.} 2020, Nature, 585,
  357, \dodoi{10.1038/s41586-020-2649-2}

\bibitem[{{Helled}(2023)}]{Helled2023}
{Helled}, R. 2023, \aap, 675, L8, \dodoi{10.1051/0004-6361/202346850}

\bibitem[{{Helled} {et~al.}(2020){Helled}, {Mazzola}, \&
  {Redmer}}]{Helled_2020b}
{Helled}, R., {Mazzola}, G., \& {Redmer}, R. 2020, Nature Reviews Physics, 2,
  562, \dodoi{10.1038/s42254-020-0223-3}

\bibitem[{{Helled} \& {Stevenson}(2017)}]{Helled2017}
{Helled}, R., \& {Stevenson}, D. 2017, \apjl, 840, L4,
  \dodoi{10.3847/2041-8213/aa6d08}

\bibitem[{{Helled} {et~al.}(2022{\natexlab{a}}){Helled}, {Werner}, {Dorn},
  {Guillot}, {Ikoma}, {Ito}, {Kama}, {Lichtenberg}, {Miguel}, {Shorttle},
  {Tackley}, {Valencia}, \& {Vazan}}]{Helled2022_Ariel}
{Helled}, R., {Werner}, S., {Dorn}, C., {et~al.} 2022{\natexlab{a}},
  Experimental Astronomy, 53, 323, \dodoi{10.1007/s10686-021-09739-3}

\bibitem[{{Helled} {et~al.}(2022{\natexlab{b}}){Helled}, {Stevenson}, {Lunine},
  {Bolton}, {Nettelmann}, {Atreya}, {Guillot}, {Militzer}, {Miguel}, \&
  {Hubbard}}]{Helled2022_Jupiter}
{Helled}, R., {Stevenson}, D.~J., {Lunine}, J.~I., {et~al.} 2022{\natexlab{b}},
  \icarus, 378, 114937, \dodoi{10.1016/j.icarus.2022.114937}

\bibitem[{{Howard} {et~al.}(2023){Howard}, {Guillot}, {Markham}, {Helled},
  {M{\"u}ller}, {Stevenson}, {Lunine}, {Miguel}, \&
  {Nettelmann}}]{2023A&A...680L...2H}
{Howard}, S., {Guillot}, T., {Markham}, S., {et~al.} 2023, \aap, 680, L2,
  \dodoi{10.1051/0004-6361/202348129}

\bibitem[{{Idini} \& {Stevenson}(2022)}]{Idini_2022}
{Idini}, B., \& {Stevenson}, D.~J. 2022, \psj, 3, 89,
  \dodoi{10.3847/PSJ/ac6179}

\bibitem[{Jermyn {et~al.}(2023)Jermyn, Bauer, Schwab, Farmer, Ball, Bellinger,
  Dotter, Joyce, Marchant, Mombarg, Wolf, Sunny~Wong, Cinquegrana, Farrell,
  Smolec, Thoul, Cantiello, Herwig, Toloza, Bildsten, Townsend, \&
  Timmes}]{Jermyn_2023}
Jermyn, A.~S., Bauer, E.~B., Schwab, J., {et~al.} 2023, The Astrophysical
  Journal Supplement Series, 265, 15, \dodoi{10.3847/1538-4365/acae8d}

\bibitem[{{Kato}(1966)}]{Kato_1966}
{Kato}, S. 1966, \pasj, 18, 374

\bibitem[{Knierim(2024{\natexlab{a}})}]{mesa_custom_eos}
Knierim, H. 2024{\natexlab{a}}, Henrik-Knierim/mesa\_custom\_EoS: Version
  1.0.0, 1.0.0,  Zenodo, \dodoi{10.5281/zenodo.13897571}

\bibitem[{Knierim(2024{\natexlab{b}})}]{py_mesa_helper}
---. 2024{\natexlab{b}}, Henrik-Knierim/py\_mesa\_helper: Version 1.0.0, 1.0.0,
   Zenodo, \dodoi{10.5281/zenodo.13897562}

\bibitem[{Knierim {et~al.}(2022)Knierim, Shibata, \& Helled}]{Knierim_2022}
Knierim, H., Shibata, S., \& Helled, R. 2022, Astronomy {\&} Astrophysics, 665,
  L5, \dodoi{10.1051/0004-6361/202244516}

\bibitem[{{Landau} \& {Lifshitz}(1959)}]{Landau_1959}
{Landau}, L.~D., \& {Lifshitz}, E.~M. 1959, {Fluid Mechanics} (Oxford:
  Pergamon)

\bibitem[{{Laughlin} {et~al.}(2011){Laughlin}, {Crismani}, \&
  {Adams}}]{Laughlin_2011}
{Laughlin}, G., {Crismani}, M., \& {Adams}, F.~C. 2011, \apjl, 729, L7,
  \dodoi{10.1088/2041-8205/729/1/L7}

\bibitem[{{Ledoux}(1947)}]{Ledoux_1947}
{Ledoux}, P. 1947, \apj, 105, 305, \dodoi{10.1086/144905}

\bibitem[{{Lodders}(2021)}]{Lodders_2021}
{Lodders}, K. 2021, \ssr, 217, 44, \dodoi{10.1007/s11214-021-00825-8}

\bibitem[{{Lozovsky} {et~al.}(2017){Lozovsky}, {Helled}, {Rosenberg}, \&
  {Bodenheimer}}]{Lozovsky2017}
{Lozovsky}, M., {Helled}, R., {Rosenberg}, E.~D., \& {Bodenheimer}, P. 2017,
  \apj, 836, 227, \dodoi{10.3847/1538-4357/836/2/227}

\bibitem[{{Madhusudhan}(2012)}]{Madhusudhan_2012}
{Madhusudhan}, N. 2012, \apj, 758, 36, \dodoi{10.1088/0004-637X/758/1/36}

\bibitem[{Madhusudhan(2019)}]{Madhusudhan_2019}
Madhusudhan, N. 2019, Annual Review of Astronomy and Astrophysics, 57, 617,
  \dodoi{https://doi.org/10.1146/annurev-astro-081817-051846}

\bibitem[{Madhusudhan {et~al.}(2017)Madhusudhan, Bitsch, Johansen, \&
  Eriksson}]{Madhusudhan_2017}
Madhusudhan, N., Bitsch, B., Johansen, A., \& Eriksson, L. 2017, Monthly
  Notices of the Royal Astronomical Society, 469, 4102,
  \dodoi{10.1093/mnras/stx1139}

\bibitem[{{Mankovich} \& {Fuller}(2021)}]{mankovich2021}
{Mankovich}, C.~R., \& {Fuller}, J. 2021, Nature Astronomy, 5, 1103,
  \dodoi{10.1038/s41550-021-01448-3}

\bibitem[{{Marley} {et~al.}(2007){Marley}, {Fortney}, {Hubickyj},
  {Bodenheimer}, \& {Lissauer}}]{Marley2007}
{Marley}, M.~S., {Fortney}, J.~J., {Hubickyj}, O., {Bodenheimer}, P., \&
  {Lissauer}, J.~J. 2007, \apj, 655, 541, \dodoi{10.1086/509759}

\bibitem[{{Miguel} {et~al.}(2022){Miguel}, {Bazot}, {Guillot}, {Howard},
  {Galanti}, {Kaspi}, {Hubbard}, {Militzer}, {Helled}, {Atreya}, {Connerney},
  {Durante}, {Kulowski}, {Lunine}, {Stevenson}, \& {Bolton}}]{Miguel2022}
{Miguel}, Y., {Bazot}, M., {Guillot}, T., {et~al.} 2022, \aap, 662, A18,
  \dodoi{10.1051/0004-6361/202243207}

\bibitem[{{Molli{\`e}re} {et~al.}(2022){Molli{\`e}re}, {Molyarova}, {Bitsch},
  {Henning}, {Schneider}, {Kreidberg}, {Eistrup}, {Burn}, {Nasedkin},
  {Semenov}, {Mordasini}, {Schlecker}, {Schwarz}, {Lacour}, {Nowak}, \&
  {Schulik}}]{Molliere_2022}
{Molli{\`e}re}, P., {Molyarova}, T., {Bitsch}, B., {et~al.} 2022, \apj, 934,
  74, \dodoi{10.3847/1538-4357/ac6a56}

\bibitem[{{Mordasini}(2013)}]{Mordasini2013}
{Mordasini}, C. 2013, \aap, 558, A113, \dodoi{10.1051/0004-6361/201321617}

\bibitem[{{Mordasini} {et~al.}(2017){Mordasini}, {Marleau}, \&
  {Molli{\`e}re}}]{Mordasini2017}
{Mordasini}, C., {Marleau}, G.~D., \& {Molli{\`e}re}, P. 2017, \aap, 608, A72,
  \dodoi{10.1051/0004-6361/201630077}

\bibitem[{{M{\"u}ller} {et~al.}(2020{\natexlab{a}}){M{\"u}ller}, {Ben-Yami}, \&
  {Helled}}]{Mueller_2020b}
{M{\"u}ller}, S., {Ben-Yami}, M., \& {Helled}, R. 2020{\natexlab{a}}, \apj,
  903, 147, \dodoi{10.3847/1538-4357/abba19}

\bibitem[{M{\"u}ller \& Helled(2021)}]{Mueller_2021}
M{\"u}ller, S., \& Helled, R. 2021, Monthly Notices of the Royal Astronomical
  Society, 507, 2094, \dodoi{10.1093/mnras/stab2250}

\bibitem[{{M{\"u}ller} \& {Helled}(2023{\natexlab{a}})}]{Muller2023}
{M{\"u}ller}, S., \& {Helled}, R. 2023{\natexlab{a}}, Frontiers in Astronomy
  and Space Sciences, 10, 1179000, \dodoi{10.3389/fspas.2023.1179000}

\bibitem[{{M{\"u}ller} \& {Helled}(2023{\natexlab{b}})}]{Mueller_2023}
---. 2023{\natexlab{b}}, \aap, 669, A24, \dodoi{10.1051/0004-6361/202244827}

\bibitem[{{M{\"u}ller} {et~al.}(2020{\natexlab{b}}){M{\"u}ller}, {Helled}, \&
  {Cumming}}]{Mueller2020a}
{M{\"u}ller}, S., {Helled}, R., \& {Cumming}, A. 2020{\natexlab{b}}, \aap, 638,
  A121, \dodoi{10.1051/0004-6361/201937376}

\bibitem[{M{\"u}ller {et~al.}(2020)M{\"u}ller, Helled, \&
  Cumming}]{Mueller_2020}
M{\"u}ller, S., Helled, R., \& Cumming, A. 2020, Astronomy {\&} Astrophysics
  Astrophysics, 638, A121, \dodoi{10.1051/0004-6361/201937376}

\bibitem[{{Nettelmann} {et~al.}(2021){Nettelmann}, {Movshovitz}, {Ni},
  {Fortney}, {Galanti}, {Kaspi}, {Helled}, {Mankovich}, \&
  {Bolton}}]{2021PSJ.....2..241N}
{Nettelmann}, N., {Movshovitz}, N., {Ni}, D., {et~al.} 2021, \psj, 2, 241,
  \dodoi{10.3847/PSJ/ac390a}

\bibitem[{{\"O}berg {et~al.}(2011){\"O}berg, Murray-Clay, \&
  Bergin}]{Oeberg_2011}
{\"O}berg, K.~I., Murray-Clay, R., \& Bergin, E.~A. 2011, The Astrophysical
  Journal, 743, L16, \dodoi{10.1088/2041-8205/743/1/l16}

\bibitem[{Parmentier {et~al.}(2015)Parmentier, Guillot, Fortney, \&
  Marley}]{Parmentier_2015}
Parmentier, V., Guillot, T., Fortney, J.~J., \& Marley, M.~S. 2015, Astronomy
  \& Astrophysics, 574, A35

\bibitem[{Paxton {et~al.}(2010)Paxton, Bildsten, Dotter, Herwig, Lesaffre, \&
  Timmes}]{Paxton_2011}
Paxton, B., Bildsten, L., Dotter, A., {et~al.} 2010, The Astrophysical Journal
  Supplement Series, 192, 3, \dodoi{10.1088/0067-0049/192/1/3}

\bibitem[{Paxton {et~al.}(2013)Paxton, Cantiello, Arras, Bildsten, Brown,
  Dotter, Mankovich, Montgomery, Stello, Timmes, \& Townsend}]{Paxton_2013}
Paxton, B., Cantiello, M., Arras, P., {et~al.} 2013, The Astrophysical Journal
  Supplement Series, 208, 4, \dodoi{10.1088/0067-0049/208/1/4}

\bibitem[{Paxton {et~al.}(2015)Paxton, Marchant, Schwab, Bauer, Bildsten,
  Cantiello, Dessart, Farmer, Hu, Langer, Townsend, Townsley, \&
  Timmes}]{Paxton_2015}
Paxton, B., Marchant, P., Schwab, J., {et~al.} 2015, The Astrophysical Journal
  Supplement Series, 220, 15, \dodoi{10.1088/0067-0049/220/1/15}

\bibitem[{Paxton {et~al.}(2018)Paxton, Schwab, Bauer, Bildsten, Blinnikov,
  Duffell, Farmer, Goldberg, Marchant, Sorokina, Thoul, Townsend, \&
  Timmes}]{Paxton_2018}
Paxton, B., Schwab, J., Bauer, E.~B., {et~al.} 2018, The Astrophysical Journal
  Supplement Series, 234, 34, \dodoi{10.3847/1538-4365/aaa5a8}

\bibitem[{Paxton {et~al.}(2019)Paxton, Smolec, Schwab, Gautschy, Bildsten,
  Cantiello, Dotter, Farmer, Goldberg, Jermyn, Kanbur, Marchant, Thoul,
  Townsend, Wolf, Zhang, \& Timmes}]{Paxton_2019}
Paxton, B., Smolec, R., Schwab, J., {et~al.} 2019, The Astrophysical Journal
  Supplement Series, 243, 10, \dodoi{10.3847/1538-4365/ab2241}

\bibitem[{{Pollack} {et~al.}(1996){Pollack}, {Hubickyj}, {Bodenheimer},
  {Lissauer}, {Podolak}, \& {Greenzweig}}]{Pollack1996}
{Pollack}, J.~B., {Hubickyj}, O., {Bodenheimer}, P., {et~al.} 1996, \icarus,
  124, 62, \dodoi{10.1006/icar.1996.0190}

\bibitem[{{Radko}(2007)}]{Radko_2007}
{Radko}, T. 2007, Journal of Fluid Mechanics, 577, 251,
  \dodoi{10.1017/S0022112007004703}

\bibitem[{{Rosenblum} {et~al.}(2011){Rosenblum}, {Garaud}, {Traxler}, \&
  {Stellmach}}]{Rosenblum_2011}
{Rosenblum}, E., {Garaud}, P., {Traxler}, A., \& {Stellmach}, S. 2011, \apj,
  731, 66, \dodoi{10.1088/0004-637X/731/1/66}

\bibitem[{{Saumon} {et~al.}(1995){Saumon}, {Chabrier}, \& {van
  Horn}}]{Saumon1995}
{Saumon}, D., {Chabrier}, G., \& {van Horn}, H.~M. 1995, \apjs, 99, 713,
  \dodoi{10.1086/192204}

\bibitem[{{Schneider} \& {Bitsch}(2021)}]{Schneider_2021b}
{Schneider}, A.~D., \& {Bitsch}, B. 2021, \aap, 654, A72,
  \dodoi{10.1051/0004-6361/202141096}

\bibitem[{{Stevenson} {et~al.}(2022){Stevenson}, {Bodenheimer}, {Lissauer}, \&
  {D'Angelo}}]{Stevenson2022}
{Stevenson}, D.~J., {Bodenheimer}, P., {Lissauer}, J.~J., \& {D'Angelo}, G.
  2022, \psj, 3, 74, \dodoi{10.3847/PSJ/ac5c44}

\bibitem[{{Stevenson} \& {Salpeter}(1977)}]{Stevenson_1977b}
{Stevenson}, D.~J., \& {Salpeter}, E.~E. 1977, \apjs, 35, 239,
  \dodoi{10.1086/190479}

\bibitem[{{Sur} {et~al.}(2024){Sur}, {Su}, {Tejada Arevalo}, {Chen}, \&
  {Burrows}}]{Sur_2024}
{Sur}, A., {Su}, Y., {Tejada Arevalo}, R., {Chen}, Y.-X., \& {Burrows}, A.
  2024, \apj, 971, 104, \dodoi{10.3847/1538-4357/ad57c3}

\bibitem[{{Tinetti} {et~al.}(2018){Tinetti}, {Drossart}, {Eccleston},
  {Hartogh}, {Heske}, {Leconte}, {Micela}, {Ollivier}, {Pilbratt}, {Puig},
  {Turrini}, {Vandenbussche}, {Wolkenberg}, {Beaulieu}, {Buchave}, {Ferus},
  {Griffin}, {Guedel}, {Justtanont}, {Lagage}, {Machado}, {Malaguti}, {Min},
  {N{\o}rgaard-Nielsen}, {Rataj}, {Ray}, {Ribas}, {Swain}, {Szabo}, {Werner},
  {Barstow}, {Burleigh}, {Cho}, {du Foresto}, {Coustenis}, {Decin}, {Encrenaz},
  {Galand }, {Gillon}, {Helled}, {Morales}, {Mu{\~n}oz}, {Moneti}, {Pagano},
  {Pascale}, {Piccioni}, {Pinfield}, {Sarkar}, {Selsis}, {Tennyson}, {Triaud},
  {Venot}, {Waldmann}, {Waltham}, {Wright}, {Amiaux}, {Augu{\`e}res},
  {Berth{\'e}}, {Bezawada}, {Bishop}, {Bowles}, {Coffey}, {Colom{\'e}},
  {Crook}, {Crouzet}, {Da Peppo}, {Sanz}, {Focardi}, {Frericks}, {Hunt},
  {Kohley}, {Middleton}, {Morgante}, {Ottensamer}, {Pace}, {Pearson},
  {Stamper}, {Symonds}, {Rengel}, {Renotte}, {Ade}, {Affer}, {Alard}, {Allard},
  {Altieri}, {Andr{\'e}}, {Arena}, {Argyriou}, {Aylward}, {Baccani}, {Bakos},
  {Banaszkiewicz}, {Barlow}, {Batista}, {Bellucci}, {Benatti}, {Bernardi},
  {B{\'e}zard}, {Blecka}, {Bolmont}, {Bonfond}, {Bonito}, {Bonomo}, {Brucato},
  {Brun}, {Bryson}, {Bujwan}, {Casewell}, {Charnay}, {Pestellini}, {Chen},
  {Ciaravella}, {Claudi}, {Cl{\'e}dassou}, {Damasso}, {Damiano}, {Danielski},
  {Deroo}, {Di Giorgio}, {Dominik}, {Doublier}, {Doyle}, {Doyon}, {Drummond},
  {Duong}, {Eales}, {Edwards}, {Farina}, {Flaccomio}, {Fletcher}, {Forget},
  {Fossey}, {Fr{\"a}nz}, {Fujii}, {Garc{\'\i}a-Piquer}, {Gear}, {Geoffray},
  {G{\'e}rard}, {Gesa}, {Gomez}, {Graczyk}, {Griffith}, {Grodent}, {Guarcello},
  {Gustin}, {Hamano}, {Hargrave}, {Hello}, {Heng}, {Herrero}, {Hornstrup},
  {Hubert}, {Ida}, {Ikoma}, {Iro}, {Irwin}, {Jarchow}, {Jaubert}, {Jones},
  {Julien}, {Kameda}, {Kerschbaum}, {Kervella}, {Koskinen}, {Krijger}, {Krupp},
  {Lafarga}, {Landini}, {Lellouch}, {Leto}, {Luntzer}, {Rank-L{\"u}ftinger},
  {Maggio}, {Maldonado}, {Maillard}, {Mall}, {Marquette}, {Mathis}, {Maxted},
  {Matsuo}, {Medvedev}, {Miguel}, {Minier}, {Morello}, {Mura}, {Narita},
  {Nascimbeni}, {Nguyen Tong}, {Noce}, {Oliva}, {Palle}, {Palmer}, {Pancrazzi},
  {Papageorgiou}, {Parmentier}, {Perger}, {Petralia}, {Pezzuto},
  {Pierrehumbert}, {Pillitteri}, {Piotto}, {Pisano}, {Prisinzano}, {Radioti},
  {R{\'e}ess}, {Rezac}, {Rocchetto}, {Rosich}, {Sanna}, {Santerne}, {Savini},
  {Scandariato}, {Sicardy}, {Sierra}, {Sindoni}, {Skup}, {Snellen}, {Sobiecki},
  {Soret}, {Sozzetti}, {Stiepen}, {Strugarek}, {Taylor}, {Taylor}, {Terenzi},
  {Tessenyi}, {Tsiaras}, {Tucker}, {Valencia}, {Vasisht}, {Vazan}, {Vilardell},
  {Vinatier}, {Viti}, {Waters}, {Wawer}, {Wawrzaszek}, {Whitworth}, {Yung},
  {Yurchenko}, {Osorio}, {Zellem}, {Zingales}, \& {Zwart}}]{Tinetti2018}
{Tinetti}, G., {Drossart}, P., {Eccleston}, P., {et~al.} 2018, Experimental
  Astronomy, 46, 135, \dodoi{10.1007/s10686-018-9598-x}

\bibitem[{{Tulekeyev} {et~al.}(2024){Tulekeyev}, {Garaud}, {Idini}, \&
  {Fortney}}]{Tulekeyev_2024}
{Tulekeyev}, A., {Garaud}, P., {Idini}, B., \& {Fortney}, J.~J. 2024, \psj, 5,
  190, \dodoi{10.3847/PSJ/ad6571}

\bibitem[{{Turrini} {et~al.}(2018){Turrini}, {Miguel}, {Zingales}, {Piccialli},
  {Helled}, {Vazan}, {Oliva}, {Sindoni}, {Pani{\'c}}, {Leconte}, {Min},
  {Pirani}, {Selsis}, {Coud{\'e} du Foresto}, {Mura}, \&
  {Wolkenberg}}]{Turrini_2018}
{Turrini}, D., {Miguel}, Y., {Zingales}, T., {et~al.} 2018, Experimental
  Astronomy, 46, 45, \dodoi{10.1007/s10686-017-9570-1}

\bibitem[{{Turrini} {et~al.}(2021){Turrini}, {Schisano}, {Fonte}, {Molinari},
  {Politi}, {Fedele}, {Pani{\'c}}, {Kama}, {Changeat}, \&
  {Tinetti}}]{Turrini_2021}
{Turrini}, D., {Schisano}, E., {Fonte}, S., {et~al.} 2021, \apj, 909, 40,
  \dodoi{10.3847/1538-4357/abd6e5}

\bibitem[{{Valletta} \& {Helled}(2020)}]{Valletta2020}
{Valletta}, C., \& {Helled}, R. 2020, \apj, 900, 133,
  \dodoi{10.3847/1538-4357/aba904}

\bibitem[{{Vazan} {et~al.}(2018){Vazan}, {Helled}, \& {Guillot}}]{Vazan2018}
{Vazan}, A., {Helled}, R., \& {Guillot}, T. 2018, \aap, 610, L14,
  \dodoi{10.1051/0004-6361/201732522}

\bibitem[{Virtanen {et~al.}(2020)Virtanen, Gommers, Oliphant, Haberland, Reddy,
  Cournapeau, Burovski, Peterson, Weckesser, Bright, {van der Walt}, Brett,
  Wilson, Millman, Mayorov, Nelson, Jones, Kern, Larson, Carey, Polat, Feng,
  Moore, {VanderPlas}, Laxalde, Perktold, Cimrman, Henriksen, Quintero, Harris,
  Archibald, Ribeiro, Pedregosa, {van Mulbregt}, \& {SciPy 1.0
  Contributors}}]{2020SciPy-NMeth}
Virtanen, P., Gommers, R., Oliphant, T.~E., {et~al.} 2020, Nature Methods, 17,
  261, \dodoi{10.1038/s41592-019-0686-2}

\bibitem[{{Wahl} {et~al.}(2017){Wahl}, {Hubbard}, {Militzer}, {Guillot},
  {Miguel}, {Movshovitz}, {Kaspi}, {Helled}, {Reese}, {Galanti}, {Levin},
  {Connerney}, \& {Bolton}}]{Wahl2017}
{Wahl}, S.~M., {Hubbard}, W.~B., {Militzer}, B., {et~al.} 2017, \grl, 44, 4649,
  \dodoi{10.1002/2017GL073160}

\end{thebibliography}
\bibliographystyle{aasjournal}

\end{document}